\documentclass[12pt,reqno]{amsart}%
\usepackage{graphicx}
\usepackage{amscd}
\usepackage{amsmath}
\usepackage{epsfig}
\usepackage{amsfonts}
\usepackage{amssymb}%
\providecommand{\U}[1]{\protect\rule{.1in}{.1in}}
\providecommand{\U}[1]{\protect\rule{.1in}{.1in}}
\providecommand{\U}[1]{\protect\rule{.1in}{.1in}}
\allowdisplaybreaks

\theoremstyle{plain}

\numberwithin{equation}{section}

\begin{document}
\title[Time-Dependent Schr\"{o}dinger Equation]{Solution of the Cauchy Problem for %
a Time-Dependent Schr\"{o}dinger Equation}
\author{Maria Meiler}
\address{Department of Mathematics, Munich University of Technology, Boltzmannstrasse
3, D-85747, Garching, Munich, Germany}
\email{Maria.Meiler@web.de}
\author{Ricardo Cordero--Soto}
\address{Department of Mathematics and Statistics, Arizona State University, Tempe, AZ
85287, U.S.A.}
\email{ricardojavier81@gmail.com}
\author{Sergei K. Suslov}
\address{Department of Mathematics and Statistics, Arizona State University, Tempe, AZ
85287, U.S.A.}
\email{sks@asu.edu}
\urladdr{http://hahn.la.asu.edu/\symbol{126}suslov/index.html}
\date{\today}

\begin{abstract}
We construct an explicit solution of the Cauchy initial value problem for the
$n$-dimen\-sional Schr\"{o}dinger equation with certain time-dependent
Hamiltonian operator of a modified oscillator. The dynamical $SU\left(
1,1\right)  $ symmetry of the harmonic oscillator wave functions, Bargmann's
functions for the discrete positive series of the irreducible representations
of this group, the Fourier integral of a weighted product of the
Meixner--Pollaczek polynomials, a Hankel-type integral transform and the
hyperspherical harmonics are utilized in order to derive the corresponding
Green function. It is then generalized to a case of the forced modified
oscillator. The propagators for two models of the relativistic oscillator are
also found. An expansion formula of a plane wave in terms of the
hyperspherical harmonics and solution of certain infinite system of ordinary
differential equations are derived as a by-product.

\end{abstract}
\maketitle

\section{Introduction}

The time-dependent Schr\"{o}dinger equation for a free particle%
\begin{equation}
i\psi_{t}+\Delta\psi=0 \label{i1}%
\end{equation}
in the Euclidean space of $n$ dimensions $\boldsymbol{R}^{n}$ can be rewritten
in a Hamiltonian form as%
\begin{equation}
i\frac{\partial\psi}{\partial t}=H\psi,\qquad H=-\Delta=\frac{1}{2}\sum
_{s=1}^{n}\left(  a_{s}+a_{s}^{\dagger}\right)  ^{2}, \label{i2}%
\end{equation}
where $a_{s}^{\dagger}$ and $a_{s}$ are the creation and annihilation
operators, respectively, given by%
\begin{equation}
a_{s}^{\dagger}=\frac{1}{i\sqrt{2}}\left(  \frac{\partial}{\partial x_{s}%
}-x_{s}\right)  ,\qquad a_{s}=\frac{1}{i\sqrt{2}}\left(  \frac{\partial
}{\partial x_{s}}+x_{s}\right)  \label{i3}%
\end{equation}
as in \cite{Flu}. They satisfy the familiar commutation relations%
\begin{equation}
\left[  a_{s},a_{s^{\prime}}\right]  =\left[  a_{s}^{\dagger},a_{s^{\prime}%
}^{\dagger}\right]  =0,\qquad\left[  a_{s},a_{s^{\prime}}^{\dagger}\right]
=\delta_{ss^{\prime}}\qquad\left(  s,s^{\prime}=1,2,...\ ,n\right)  ,
\label{i4}%
\end{equation}
which are invariant under the transformation%
\begin{equation}
a_{s}\rightarrow a_{s}\left(  t\right)  =e^{it}a_{s},\qquad a_{s}^{\dagger
}\rightarrow a_{s}^{\dagger}\left(  t\right)  =a_{s}^{\dagger}e^{-it}.
\label{i5}%
\end{equation}
The substitution%
\begin{equation}
H\rightarrow H\left(  t\right)  =\frac{1}{2}\sum_{s=1}^{n}\left(  a_{s}\left(
t\right)  +a_{s}^{\dagger}\left(  t\right)  \right)  ^{2} \label{i6}%
\end{equation}
results in the time-dependent Schr\"{o}dinger equation for a modified
oscillator%
\begin{equation}
i\frac{\partial\psi}{\partial t}=H\left(  t\right)  \psi\label{i7}%
\end{equation}
with the Hamiltonian of the form%
\begin{equation}
H\left(  t\right)  =\frac{1}{2}\sum_{s=1}^{n}\left(  a_{s}a_{s}^{\dagger
}+a_{s}^{\dagger}a_{s}\right)  +\frac{1}{2}e^{2it}\sum_{s=1}^{n}\left(
a_{s}\right)  ^{2}+\frac{1}{2}e^{-2it}\sum_{s=1}^{n}\left(  a_{s}^{\dagger
}\right)  ^{2}. \label{i8}%
\end{equation}
In the present paper we construct an exact solution of this equation subject
to the initial condition%
\begin{equation}
\left.  \psi\left(  \boldsymbol{x},t\right)  \right\vert _{t=0}=\psi
_{0}\left(  \boldsymbol{x}\right)  , \label{i9}%
\end{equation}
where $\psi_{0}\left(  \boldsymbol{x}\right)  $ in an arbitrary square
integrable complex valued function from $\mathcal{L}^{2}\left(  \boldsymbol{R}%
^{n}\right)  .$ The explicit form of equation (\ref{i7}) is given by
(\ref{sol0a}) below.\smallskip\

The paper is organized as follows. In section~2 we remind the reader about the
solution of the stationary Schr\"{o}dinger equation for the $n$-dimentional
harmonic oscillator in hyperspherical coordinates and discuss the
corresponding dynamical $SU\left(  1,1\right)  $ symmetry group. In section~3
we consider the eigenfunction expansion for the time-dependent Schr\"{o}dinger
equation (\ref{i7}). The series solution of the initial value problem
(\ref{i7})--(\ref{i9}) is obtained in section~6 after discussion of the
Meixner--Pollaczek polynomials and evaluation of the Fourier integral of their
weighted product in sections~4 and 5 respectively. Finally we construct the
corresponding Green function in section~8, while introducing a required
integral transform in section~7. A generalization to the case of the forced
modified oscillator is given in section~9. An expansion formula for the plane
wave in terms of the hyperspherical harmonics in $\boldsymbol{R}^{n}$ and its
special cases are discussed in section~10. A certain type of the
time-dependent Schr\"{o}dinger equation is considered in section~11 in an
abstract form. The propagators for two models of the relativistic oscillator
are derived in the next two sections. An axillary solution of an infinite
system of the ordinary differential equations is found in section~14 as a
by-product. Appendix at the end of the paper contains another required
integral evaluation.\smallskip

The exact solution of the one-dimensional time-dependent Schr\"{o}dinger
equation for a forced harmonic oscillator is constructed in \cite{FeynmanPhD},
\cite{Feynman}, \cite{Fey:Hib}, and \cite{Lop:Sus}; see also references
therein. These simple exactly solvable models may be of interest in a general
treatment of the non-linear time-dependent Schr\"{o}dinger equation; see
\cite{Howland}, \cite{Jafaev}, \cite{Naibo:Stef}, \cite{Rod:Schlag},
\cite{Schlag}, \cite{Yajima} and references therein. They may also be useful
as test solutions for numerical methods of solving the time-dependent
Schr\"{o}dinger equation.

\section{Dynamical Symmetry of the Harmonic Oscillator in $n$-Dimensions}

Our time-dependent Hamiltonian operator (\ref{i8}) has the following structure%
\begin{equation}
H\left(  t\right)  =H_{0}+H_{1}\left(  t\right)  , \label{harm1}%
\end{equation}
where%
\begin{equation}
H_{0}=\frac{1}{2}\sum_{s=1}^{n}\left(  a_{s}a_{s}^{\dagger}+a_{s}^{\dagger
}a_{s}\right)  \label{harm2}%
\end{equation}
is the Hamiltonian of the $n$-dimensional harmonic oscillator and%
\begin{equation}
H_{1}\left(  t\right)  =\frac{1}{2}e^{2it}\sum_{s=1}^{n}\left(  a_{s}\right)
^{2}+\frac{1}{2}e^{-2it}\sum_{s=1}^{n}\left(  a_{s}^{\dagger}\right)  ^{2}
\label{harm3}%
\end{equation}
is the part depending on time $t.$\smallskip

The stationary Schr\"{o}dinger equation for the harmonic oscillator in
$n$-dimensions%
\begin{equation}
H_{0}\Psi=E\Psi,\qquad H_{0}=\frac{1}{2}\sum_{s=1}^{n}\left(  -\frac
{\partial^{2}}{\partial x_{s}^{2}}+x_{s}^{2}\right)  \label{harm4}%
\end{equation}
can be solved explicitly in Cartesian and (hyper)spherical coordinate systems;
see, for example, \cite{Ni:Su:Uv}, \cite{Smir:Shit} and references
therein.\smallskip

In spherical coordinates $r,\Omega$ given by a certain binary tree $T,$ see
\cite{Ni:Su:Uv}, \cite{Smir:Shit}, \cite{Vil} and references therein for a
graphical approach of Vilenkin, Kuznetsov and Smorodinski\u{\i} to the theory
of (hyper)spherical harmonics, we look for solution in the form%
\begin{equation}
\psi=Y_{K\nu}\left(  \Omega\right)  \ R\left(  r\right)  , \label{harm5}%
\end{equation}
where $Y_{K\nu}$ are the spherical harmonics constructed by the given tree
$T,$ the integer number $K$ corresponds to the constant of separation of the
variables at the root of $T$ (denoted by $K$ due to the tradition of the
method of $K$-harmonics in nuclear physics \cite{Smir:Shit}) and $\nu=\left\{
l_{1},l_{2},...\ ,l_{p}\right\}  $ is the set of all other subscripts
corresponding to the remaining vertexes of the binary tree $T.$ The radial
wave function $R\left(  r\right)  ,$ which satisfies the normalization
condition%
\begin{equation}
\int_{0}^{\infty}R^{2}\left(  r\right)  \ r^{n-1}dr=1, \label{harm6}%
\end{equation}
is as follows%
\begin{equation}
R=R_{NK}\left(  r\right)  =\sqrt{\frac{2\left[  \left(  N-K\right)  /2\right]
!}{\Gamma\left[  \left(  N+K+n\right)  /2\right]  }}\ \exp\left(
-r^{2}/2\right)  \ r^{K}\ L_{\left(  N-K\right)  /2}^{K+n/2-1}\left(
r^{2}\right)  , \label{harm7}%
\end{equation}
where $L_{k}^{\alpha}\left(  \xi\right)  $ are the Laguerre polynomials; see
\cite{An:As:Ro}, \cite{Askey}, \cite{As:Wi}, \cite{Chihara}, \cite{Erd},
\cite{Ga:Ra}, \cite{Ko:Sw}, \cite{Ni:Su:Uv}, \cite{Ni:Uv}, \cite{Sze},
\cite{Vil}, \cite{Wil}, and references therein for the advanced theory of the
classical orthogonal polynomials.\smallskip

The corresponding energy levels are%
\begin{equation}
E=N+n/2,\qquad\left(  N-K\right)  /2=k=0,1,2,...\ \label{harm8}%
\end{equation}
and the normalized wave functions are given by%
\begin{equation}
\Psi=\Psi_{NK\nu}\left(  r,\Omega\right)  =Y_{K\nu}\left(  \Omega\right)
\ R_{NK}\left(  r\right)  , \label{harm9}%
\end{equation}
where $Y_{K\nu}\left(  \Omega\right)  $ are the spherical harmonics associated
with the tree $T$ and the radial functions $R_{NK}\left(  r\right)  $ are
defined by (\ref{harm7}). The wave functions of the one-dimensional harmonic
oscillator%
\begin{equation}
\Psi_{N}\left(  x\right)  =\frac{1}{\sqrt{2^{N}N!\sqrt{\pi}}}\ e^{-x^{2}%
/2}H_{N}\left(  x\right)  \label{harm9a}%
\end{equation}
can be obtain from (\ref{harm9}) by letting $n=1$ and $K=0,1$ and invoking the
familiar relations%
\begin{equation}
H_{2k}\left(  \xi\right)  =\left(  -1\right)  ^{k}2^{2k}k!\ L_{k}%
^{-1/2}\left(  \xi^{2}\right)  ,\qquad H_{2k+1}\left(  \xi\right)  =\left(
-1\right)  ^{k}2^{2k+1}k!\ \xi L_{k}^{1/2}\left(  \xi^{2}\right)
\label{harm9b}%
\end{equation}
between the Laguerre $L_{k}^{\alpha}\left(  \xi\right)  $ and Hermite
$H_{k}\left(  \xi\right)  $ polynomials, respectively. Thus%
\begin{equation}
\Psi_{N}\left(  x\right)  =\left\{
\begin{array}
[c]{c}%
\dfrac{\left(  -1\right)  ^{N/2}}{\sqrt{2}}\ R_{N0}\left(  x\right)
\bigskip,\\
\dfrac{\left(  -1\right)  ^{\left(  N-1\right)  /2}}{\sqrt{2}}\ R_{N1}\left(
x\right)  ,
\end{array}
\right.  \label{harm9c}%
\end{equation}
for even and odd $N,$ respectively; see \cite{Ni:Su:Uv} and \cite{Smir:Shit}
for more details.\smallskip

The $n$-dimensional oscillator wave functions (\ref{harm9}) have the following
group-theoretical properties. Introducing operators%
\begin{align}
J_{+}  &  =\frac{1}{2}\sum_{s=1}^{n}\left(  a_{s}^{\dagger}\right)
^{2},\qquad J_{-}=\frac{1}{2}\sum_{s=1}^{n}\left(  a_{s}\right)
^{2},\label{harm10}\\
J_{0}  &  =\frac{1}{2}\sum_{s=1}^{n}\left(  a_{s}^{\dagger}a_{s}%
+a_{s}^{\dagger}a_{s}\right)  =\frac{1}{2}H_{0},\nonumber
\end{align}
one can easily verify the following commutation relations%
\begin{equation}
\left[  J_{0},J_{\pm}\right]  =\pm J_{\pm},\qquad\left[  J_{+},J_{-}\right]
=-2J_{0}. \label{harm11}%
\end{equation}
For the Hermitian operators%
\begin{equation}
J_{x}=\frac{1}{2}\left(  J_{+}+J_{-}\right)  ,\qquad J_{y}=\frac{1}{2i}\left(
J_{+}-J_{-}\right)  ,\qquad J_{z}=J_{0}, \label{harm11a}%
\end{equation}
we get%
\begin{equation}
\left[  J_{x},J_{y}\right]  =-iJ_{z},\qquad\left[  J_{y},J_{z}\right]
=iJ_{x},\qquad\left[  J_{z},J_{x}\right]  =iJ_{y}. \label{harm11b}%
\end{equation}
These commutation rules are valid for the infinitesimal operators of the
non-compact group $SU\left(  1,1\right)  ;$ see, for example,
\cite{Bargmann47}, \cite{Fil:Ovch:Smir}, \cite{Ni:Su:Uv} and \cite{Smir:Shit}
for more details.\smallskip

We are going to use a different notation for the wave function (\ref{harm9})
as follows%
\begin{equation}
\psi_{jm}\left(  \boldsymbol{x}\right)  =\psi_{jm\left\{  \nu\right\}
}\left(  \boldsymbol{x}\right)  =\Psi_{NK\nu}\left(  r,\Omega\right)
=Y_{K\nu}\left(  \Omega\right)  \ R_{NK}\left(  r\right)  , \label{harm12}%
\end{equation}
where the new quantum numbers are $j=K/2+n/4-1$ and $m=N/2+n/4$ with $m=j+1,$
$j+2,...\ .$ The inequality $m\geq j+1$ holds because of the quantization rule
(\ref{harm8}), which gives $N=K,$ $K+2,$ $K+4,...\ .$\smallskip

The operators $J_{\pm}$ and $J_{0}$ in the spherical coordinates $r,\Omega$
have the form%
\begin{equation}
J_{\pm}=\frac{1}{2}\left(  H_{0}-r^{2}\pm\frac{n}{2}\pm r\frac{\partial
}{\partial r}\right)  ,\qquad J_{0}=\frac{1}{2}H_{0} \label{harm13}%
\end{equation}
and their actions on the oscillator wave functions are%
\begin{equation}
J_{\pm}\psi_{jm}=\sqrt{\left(  m\mp j\right)  \left(  m\pm j\pm1\right)
}\ \psi_{j,m\pm1},\qquad J_{0}\psi_{jm}=m\psi_{jm}, \label{harm14}%
\end{equation}
whence%
\begin{equation}
J^{2}\psi_{jm}=j\left(  j+1\right)  \psi_{jm} \label{harm15}%
\end{equation}
with $J^{2}=J_{0}^{2}+J_{0}-J_{-}J_{+}=J_{0}^{2}-J_{0}-J_{+}J_{-}\ .$ These
relations coincide with the formulas that define the action of the
infinitesimal operators $J_{\pm}$ and $J_{0}$ of the group $SU\left(
1,1\right)  $ on a basis $\left\vert j,m\right\rangle $ of the irreducible
representation $\mathcal{D}_{+}^{j}$ belonging to the discrete positive series
in an abstract Hilbert space \cite{Bargmann47}. In our realization of this
basis in terms of the wave functions (\ref{harm12}), depending on the number
$n=\dim\boldsymbol{R}^{n},$ the moment $j=K/2+n/4-1$ of the group $SU\left(
1,1\right)  $ and its projection $m=N/2+n/4$ may assume integer, half-integer
and quarter-integer values. Thus the wave functions of the $n$-dimensional
harmonic oscillator form a basis of the two-valued irreducible representation
$\mathcal{D}_{+}^{j}$ for the Lie algebra of $SU\left(  1,1\right)
.$\smallskip

Let us discuss in particular the group-theoretical properties of the wave
functions (\ref{harm9a}) of the one-dimensional harmonic oscillator. In view
of the definition (\ref{i3}) of the creation and annihilation operators,%
\begin{equation}
ia\Psi_{N}=\sqrt{N}\Psi_{N-1},\qquad-ia^{\dag}\Psi_{N}=\sqrt{N+1}\Psi_{N+1},
\label{harm16}%
\end{equation}
where we have used the familiar differentiation formulas%
\begin{equation}
H_{k}^{\prime}\left(  \xi\right)  =2kH_{k-1}\left(  \xi\right)  =2\xi
H_{k}\left(  \xi\right)  -H_{k+1}\left(  \xi\right)  \label{harm17}%
\end{equation}
for the Hermite polynomials. In this case the relations (\ref{harm14}) hold
for the basis functions of the form%
\begin{equation}
\psi_{jm}=\left\{
\begin{array}
[c]{c}%
\left(  -1\right)  ^{N/2}\Psi_{N}\bigskip,\qquad\quad\ \ N=N^{+}%
=0,2,4,\ ...\ \text{for }j=-3/4,\\
\left(  -1\right)  ^{\left(  N-1\right)  /2}\Psi_{N},\qquad N=N^{-}%
=1,3,5,\ ...\ \text{for }j=-1/4,
\end{array}
\right.  \label{harm18}%
\end{equation}
where $m=N/2+1/4.$ Thus the even and odd wave functions $\Psi_{N}\left(
x\right)  $ of the one-dimensional harmonic oscillator form, respectively,
bases for the two irreducible representations $\mathcal{D}_{+}^{j}$ of the
algebra $SU\left(  1,1\right)  $ with the moments $j=-3/4$ for the even values
of $N=N^{+}$ and $j=-1/4$ for odd $N=N^{-};$ see \cite{Ni:Su:Uv} and
\cite{Smir:Shit} for more details.

\section{Eigenfunction Expansion for the Time-Dependent Schr\"{o}dinger
Equation}

In spirit of Dirac's time-dependent perturbation theory in quantum mechanics,
see \cite{Dav}, \cite{Flu}, \cite{Gottf:T-MY}, \cite{La:Lif}, \cite{Merz},
\cite{Mes}, \cite{Schiff}, we are looking for a solution of the initial values
problem (\ref{i7})--(\ref{i9}) as an infinite multiple series%
\begin{equation}
\psi=\psi\left(  \boldsymbol{x},t\right)  =\sum_{j\left\{  \nu\right\}  }%
\sum_{m=j+1}^{\infty}c_{m}\left(  t\right)  \ \psi_{jm\left\{  \nu\right\}
}\left(  \boldsymbol{x}\right)  , \label{separ1}%
\end{equation}
where $\psi_{jm\left\{  \nu\right\}  }\left(  \boldsymbol{x}\right)  $ are the
oscillator wave functions (\ref{harm12}) depending on the space coordinates
$\boldsymbol{x}$ only and $c_{m}\left(  t\right)  =c_{jm\left\{  \nu\right\}
}\left(  t\right)  $ are yet unknown time-dependent coefficients.\smallskip

The Hamiltonian (\ref{i8}) belongs to a more general type%
\begin{equation}
H\left(  t\right)  =\omega J_{0}+\delta\left(  t\right)  J_{+}+\delta^{\ast
}\left(  t\right)  J_{-} \label{separ2}%
\end{equation}
with $\delta\left(  t\right)  =e^{-i\omega t}$ and $\omega=2;$ see
(\ref{harm10}) for the definition of operators $J_{\pm}$ and $J_{0}.$ It is
convenient to proceed further with an arbitrary value of the parameter
$\omega$ and then to choose a particular value.\smallskip

Substituting expension (\ref{separ1}) into the Schr\"{o}dinger wave equation
(\ref{i7}), with the help of (\ref{separ2}), (\ref{harm14}) and the
orthogonality property of the oscillator wave functions (\ref{harm12}),
namely,%
\begin{equation}
\int_{\boldsymbol{R}^{n}}\psi_{jm\left\{  \nu\right\}  }^{\ast}\left(
\boldsymbol{x}\right)  \psi_{j^{\prime}m^{\prime}\left\{  \nu^{\prime
}\right\}  }\left(  \boldsymbol{x}\right)  \ dv=\delta_{jj^{\prime}}%
\delta_{mm^{\prime}}\left(  \delta_{\left\{  \nu\right\}  \left\{  \nu
^{\prime}\right\}  }\right)  , \label{separ5}%
\end{equation}
we obtain an infinite system of the first order ordinary differential
equations%
\begin{align}
i\frac{dc_{m}\left(  t\right)  }{dt}  &  =\omega m\ c_{m}\left(  t\right)
+\delta\left(  t\right)  \sqrt{\left(  m-j-1\right)  \left(  m+j\right)
}\ c_{m-1}\left(  t\right) \label{separ3}\\
&  \quad+\delta^{\ast}\left(  t\right)  \sqrt{\left(  m+j+1\right)  \left(
m-j\right)  }\ c_{m+1}\left(  t\right)  .\nonumber
\end{align}
\newline The following Ansatz%
\begin{equation}
c_{m}\left(  t\right)  =\sqrt{\frac{\left(  m-j-1\right)  !}{\left(
m+j\right)  !}}\ e^{-i\omega mt}u_{m}\left(  t\right)  \label{separ4}%
\end{equation}
results in%
\begin{equation}
i\frac{du_{m}}{dt}=\delta\left(  t\right)  e^{i\omega t}\left(  m+j\right)
u_{m-1}+\delta^{\ast}\left(  t\right)  e^{-i\omega t}\left(  m-j\right)
u_{m+1}. \label{separ6}%
\end{equation}
When $\delta\left(  t\right)  =$ $e^{-i\omega t}$ we obtain the system of the
first order linear equations%
\begin{equation}
i\frac{d\boldsymbol{u}}{dt}=\mathbf{A}\boldsymbol{u},\qquad\boldsymbol{u}%
\left(  0\right)  =\boldsymbol{u}^{0}, \label{separ7a}%
\end{equation}
or, explicitly,%
\begin{equation}
i\frac{du_{m}}{dt}=\left(  m+j\right)  u_{m-1}+\left(  m-j\right)
u_{m+1}\qquad\left(  m=j+1,j+2,...\ ,\infty\right)  \label{separ7}%
\end{equation}
with the Jacobi, or three diagonal, infinite matrix $\mathbf{A}$ independent
of time $t.$ The system in hand should be solved subject to the initial
conditions%
\begin{align}
u_{m}^{0}  &  =u_{m}^{0}\left(  0\right)  =\sqrt{\frac{\left(  m+j\right)
!}{\left(  m-j-1\right)  !}}\ c_{m}\left(  0\right) \label{separ8}\\
&  =\sqrt{\frac{\left(  m+j\right)  !}{\left(  m-j-1\right)  !}}%
\ \int_{\boldsymbol{R}^{n}}\psi_{jm\left\{  \nu\right\}  }^{\ast}\left(
\boldsymbol{x}\right)  \psi_{0}\left(  \boldsymbol{x}\right)  \ dv\nonumber
\end{align}
in view of (\ref{i9}), (\ref{separ1}) and (\ref{separ4}).\smallskip

The solution of the initial value problem (\ref{separ7})--(\ref{separ8}) can
be constructed as follows%
\begin{equation}
u_{m}\left(  t\right)  =\sum_{m^{\prime}}u_{m^{\prime}}^{0}\ u_{mm^{\prime}%
}\left(  t\right)  , \label{separ9}%
\end{equation}
where $u_{mm^{\prime}}\left(  t\right)  $ is a \textquotedblleft
Green\textquotedblright\ function, or particular solutions that satisfy the
simplest initial conditions%
\begin{equation}
u_{mm^{\prime}}\left(  0\right)  =\delta_{mm^{\prime}}. \label{separ10}%
\end{equation}

Thus the solution of the original initial value problem (\ref{i7})--(\ref{i9})
is given by%
\begin{equation}
\psi\left(  \boldsymbol{x},t\right)  =\sum_{j\left\{  \nu\right\}  }%
\sum_{m=j+1}^{\infty}c_{m}\left(  t\right)  \ \psi_{jm\left\{  \nu\right\}
}\left(  \boldsymbol{x}\right)  \label{separ11}%
\end{equation}
with%
\begin{align}
c_{m}\left(  t\right)   &  =\sqrt{\frac{\left(  m-j-1\right)  !}{\left(
m+j\right)  !}}\ e^{-i\omega mt}\label{separ12}\\
&  \quad\times\sum_{m^{\prime}=j+1}^{\infty}\sqrt{\frac{\left(  m^{\prime
}+j\right)  !}{\left(  m^{\prime}-j-1\right)  !}}\ u_{mm^{\prime}}\left(
t\right)  \ \int_{\boldsymbol{R}^{n}}\psi_{jm^{\prime}\left\{  \nu\right\}
}^{\ast}\left(  \boldsymbol{x}^{\prime}\right)  \psi_{0}\left(  \boldsymbol{x}%
^{\prime}\right)  \ dv^{\prime} ,\nonumber
\end{align}
where the Green function $u_{mm^{\prime}}\left(  t\right)  $ will be
constructed in this paper in terms of the so-called Bargmann function
\cite{Bargmann47}, \cite{Ni:Su:Uv}; see also (\ref{barg2}). It will be done in
section~6 after discussion of some properties of the Meixner--Pollaczek
polynomials and an integral evaluation in the next two sections.

\section{The Meixner and Pollaczek Polynomials}

The exact solution of the initial value problem (\ref{separ7a}) can be
obtained with the help of the so-called Meixner--Pollaczek polynomials. They
can be introduced in the following way.\smallskip

The Meixner polynomials \cite{Meixner34}--\cite{Meixner42}, \cite{An:As:Ro},
\cite{Chihara}, \cite{Erd}, \cite{Ni:Su:Uv} are given by%
\begin{equation}
y_{n}\left(  x\right)  =m_{n}^{\left(  \gamma,\ \mu\right)  }\left(  x\right)
=\left(  \gamma\right)  _{n}\ _{2}F_{1}\left(
\begin{array}
[c]{c}%
-n\medskip,\ -x\\
\gamma
\end{array}
;\ 1-\frac{1}{\mu}\right)  , \label{meixner1}%
\end{equation}
where $\left(  \gamma\right)  _{n}=\gamma\left(  \gamma+1\right)
\cdot...\cdot\left(  \gamma+n-1\right)  =\Gamma\left(  \gamma+n\right)
/\Gamma\left(  \gamma\right)  ;$ see also \cite{Ba} for the definition of the
generalized hypergeometric series. Their orthogonality property is%
\begin{equation}
\sum_{k=0}^{\infty}m_{n}^{\left(  \gamma,\ \mu\right)  }\left(  k\right)
\ m_{l}^{\left(  \gamma,\ \mu\right)  }\left(  k\right)  \ \frac{\mu
^{k}\left(  \gamma\right)  _{k}}{k!}=\frac{n!\left(  \gamma\right)  _{n}}%
{\mu^{n}\left(  1-\mu\right)  ^{\gamma}}\ \delta_{nl} \label{meixner2}%
\end{equation}
with $\gamma>0$ and $0<\mu<1;$ the proof is given, for example, in
\cite{Ni:Su:Uv} and \cite{Ni:Uv}.\smallskip

An important generating relation for the hypergeometric function
\begin{align}
&  \sum_{k=0}^{\infty}\frac{\left(  r-k+1\right)  _{k}}{k!}\ s^{k}\ _{2}%
F_{1}\left(
\begin{array}
[c]{c}%
-k\medskip,\ -p\\
-r
\end{array}
;\ u\right)  \ _{2}F_{1}\left(
\begin{array}
[c]{c}%
-k\medskip,\ -q\\
-r
\end{array}
;\ v\right) \nonumber\\
&  \qquad=\left(  1+s\right)  ^{r-p-q}\left(  1+s-su\right)  ^{p}\left(
1+s-sv\right)  ^{q}\nonumber\\
&  \qquad\quad\times\ _{2}F_{1}\left(
\begin{array}
[c]{c}%
-p\medskip,\ -q\\
-r
\end{array}
;-\ \frac{suv}{\left(  1+s-su\right)  \left(  1+s-sv\right)  }\right)  ,
\label{bilinearmeixner}%
\end{align}
which is due to Meixner \cite{Meixner42}, gives an extension of the
orthogonality property as%
\begin{align}
&  \sum_{k=0}^{\infty}m_{n}^{\left(  \gamma,\ \mu\right)  }\left(  k\right)
\ m_{l}^{\left(  \gamma,\ \mu\right)  }\left(  k\right)  \ \frac{\left(  \mu
t\right)  ^{k}\left(  \gamma\right)  _{k}}{k!}\label{meixner3}\\
&  \ \ \ =\left(  \gamma\right)  _{n}\left(  \gamma\right)  _{l}%
\ \frac{\left(  1-t\right)  ^{n+l}}{\left(  1-\mu t\right)  ^{n+l+\gamma}%
}\ \ _{2}F_{1}\left(
\begin{array}
[c]{c}%
-n,\ -l\\
\gamma
\end{array}
;\ \frac{\left(  1-\mu\right)  ^{2}t}{\mu\left(  1-t\right)  ^{2}}\right)
,\quad\left\vert t\right\vert \,<1,\nonumber
\end{align}
equation~(\ref{meixner2}) arises in the limit $t\rightarrow1^{-},$ and the
explicit representation for the Poisson kernel%
\begin{align}
&  \sum_{n=0}^{\infty}m_{n}^{\left(  \gamma,\ \mu\right)  }\left(  x\right)
\ m_{n}^{\left(  \gamma,\ \mu\right)  }\left(  y\right)  \ \frac{\left(  \mu
t\right)  ^{n}}{\left(  \gamma\right)  _{n}n!}\label{meixner4}\\
&  \quad=\frac{\left(  1-t\right)  ^{x+y}}{\left(  1-\mu t\right)
^{x+y+\gamma}}\ \ _{2}F_{1}\left(
\begin{array}
[c]{c}%
-x,\ -y\\
\gamma
\end{array}
;\ \frac{\left(  1-\mu\right)  ^{2}t}{\mu\left(  1-t\right)  ^{2}}\right)
,\quad\left\vert t\right\vert \,<1.\nonumber
\end{align}
The last two equations are related to each other in view of the self-duality%
\begin{equation}
m_{n}^{\left(  \gamma,\ \mu\right)  }\left(  k\right)  /\left(  \gamma\right)
_{n}\ =m_{k}^{\left(  \gamma,\ \mu\right)  }\left(  n\right)  /\left(
\gamma\right)  _{k}\ \qquad\left(  n,k=0,1,2,...\ \right)  \label{meixner5}%
\end{equation}
of the Meixner polynomials; cf.~(\ref{meixner1}).\smallskip

The Meixner--Pollaczek polynomials \cite{Meixner34}, \cite{Pollaczek50},
\cite{Chihara}, \cite{At:SusCH}, \cite{Ni:Su:Uv} given by%
\begin{equation}
p_{n}\left(  x\right)  =P_{n}^{\lambda}\left(  x,\varphi\right)
=\frac{e^{-in\varphi}}{n!}\ m_{n}^{\left(  2\lambda,\ \mu\right)  }\left(
ix-\lambda\right)  ,\qquad\mu=e^{-2i\varphi} \label{pollaczek1}%
\end{equation}
with $\lambda>0$ and $0<\varphi<\pi$ satisfy the following three term
recurrence relation%
\begin{equation}
xP_{n}^{\lambda}\left(  x,\varphi\right)  =\frac{n+1}{2\sin\varphi}%
\ P_{n+1}^{\lambda}\left(  x,\varphi\right)  -\left(  \lambda+n\right)
\frac{\cos\varphi}{\sin\varphi}\ P_{n}^{\lambda}\left(  x,\varphi\right)
+\frac{2\lambda+n-1}{2\sin\varphi}\ P_{n-1}^{\lambda}\left(  x,\varphi\right)
\label{pollaczek2}%
\end{equation}
and the continuous orthogonality relation%
\begin{equation}
\int_{-\infty}^{\infty}P_{n}^{\lambda}\left(  x,\varphi\right)  \ P_{m}%
^{\lambda}\left(  x,\varphi\right)  \ \rho\left(  x\right)  dx=\frac
{\Gamma\left(  2\lambda+n\right)  }{n!}\ \delta_{nm} \label{pollaczek3}%
\end{equation}
with respect to the weight function%
\begin{equation}
\rho\left(  x\right)  =\frac{1}{2\pi}\left(  2\sin\varphi\right)  ^{2\lambda
}\left\vert \Gamma\left(  \lambda+ix\right)  \right\vert ^{2}e^{\left(
2\varphi-\pi\right)  x}. \label{pollaczek4}%
\end{equation}
The Poisson kernel%
\begin{align}
&  \sum_{n=0}^{\infty}\frac{n!}{\left(  2\lambda\right)  _{n}}t^{n}%
\ P_{n}^{\lambda}\left(  x,\varphi\right)  \ P_{n}^{\lambda}\left(
y,\varphi\right)  =\left(  1-t\right)  ^{-2\lambda}\left(  \frac
{1-t}{1-e^{-2i\varphi}t}\right)  ^{i\left(  x+y\right)  }\label{pollaczek5}\\
&  \qquad\qquad\times\ \ _{2}F_{1}\left(
\begin{array}
[c]{c}%
\lambda-ix,\ \lambda-iy\\
2\lambda
\end{array}
;\ -\frac{4t\sin^{2}\varphi}{\left(  1-t\right)  ^{2}}\right)  ,\quad
\left\vert t\right\vert \,<1\nonumber
\end{align}
follows directly from (\ref{meixner4}) and (\ref{pollaczek1}). See
\cite{Rahman} for a more general nonsymmetric form of this Poisson kernel; its
$q$-expensions are given in \cite{As:Ra:Su} and \cite{Ra:Su}. An extension of
the orthogonality property is the following Fourier integral%
\begin{align}
&  \frac{1}{2\pi}\int_{-\infty}^{\infty}e^{-2ixt}P_{n}^{\lambda}\left(
x,\varphi\right)  \ P_{m}^{\lambda}\left(  x,\varphi\right)  \ e^{\left(
2\varphi-\pi\right)  x}\left\vert \Gamma\left(  \lambda+ix\right)  \right\vert
^{2}dx\label{pollaczek6}\\
&  \qquad=\frac{\Gamma\left(  2\lambda+n\right)  \Gamma\left(  2\lambda
+m\right)  }{4^{\lambda}\Gamma\left(  2\lambda\right)  n!m!}\frac
{e^{i\pi\lambda}\left(  \sinh t\right)  ^{n+m}}{\left(  \cos\varphi\sinh
t+i\sin\varphi\cosh t\right)  ^{n+m+2\lambda}}\nonumber\\
&  \qquad\qquad\times\ \ _{2}F_{1}\left(
\begin{array}
[c]{c}%
-n,\ -m\\
2\lambda
\end{array}
;\ -\left(  \frac{\sin\varphi}{\sinh t}\right)  ^{2}\right)  .\nonumber
\end{align}
This integral evaluation will be given in the next section. In the limit
$t\rightarrow0$ we obtain the orthogonality relation (\ref{pollaczek3}%
)--(\ref{pollaczek4}). See also \cite{AskeyCH}, \cite{As:WiCH},
\cite{At:SusCH}, and \cite{KoelinkCH} for the introduction and properties of
the continuous Hahn polynomials that generalize the Meixner--Pollaczek polynomials.

\section{Evaluation of the Integral}

We consider the analytic continuation of the relation (\ref{meixner3}) in the
parameter $\mu.$
%%%%%%%%%%%%%%%%%%%%%%%%%%%%%%%%%%%%%%%%%%%
%%%This WinEdt version of the figure 1%%%
\begin{figure}[ptbh]
\centering\scalebox{.65}{\includegraphics{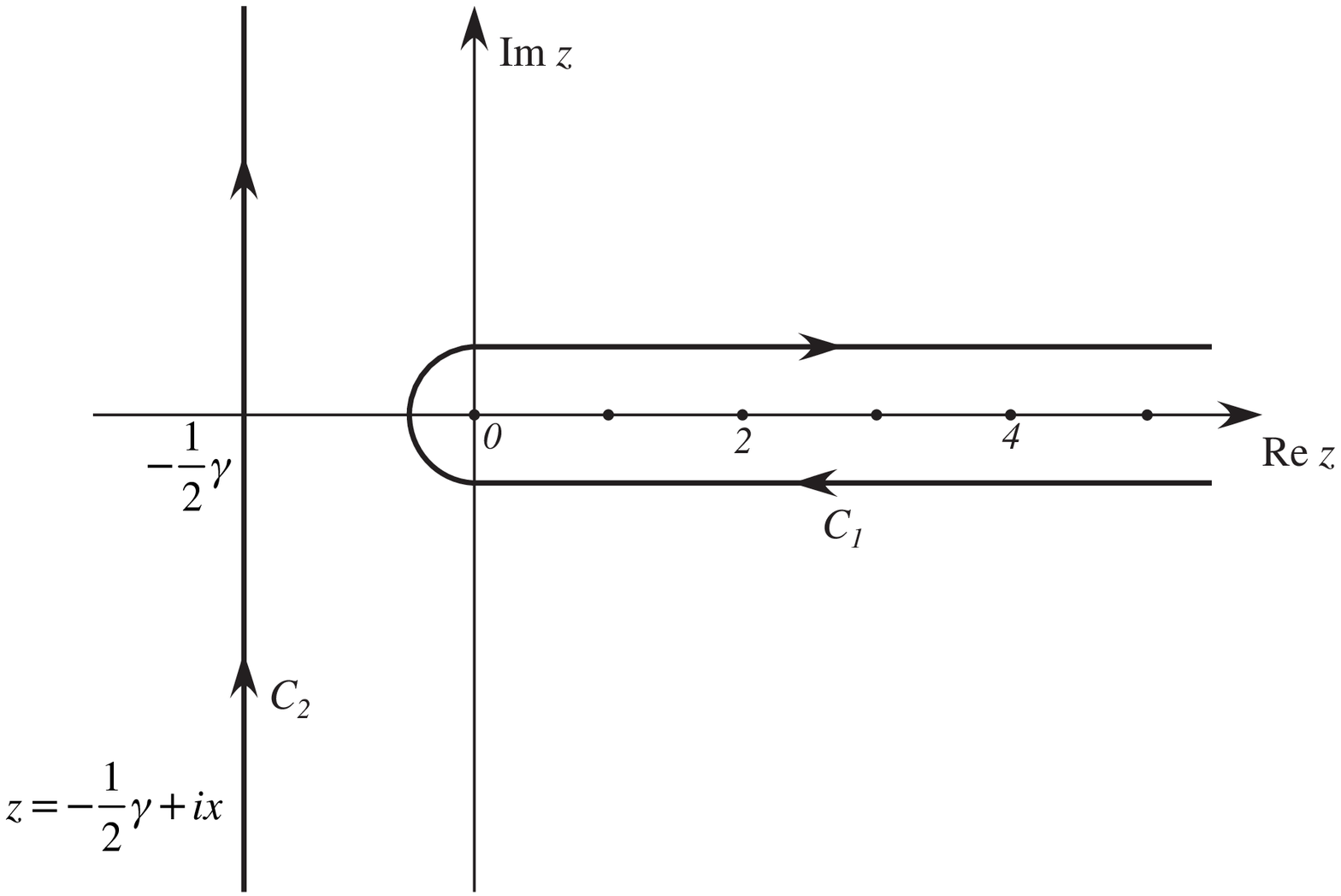}}\caption{Contours in
the complex plane.}%
\end{figure}
%%%%%%%%%%%%%%%%%%%%%%%%%%%%%%%%%%%%%%%%%%%
By the Cauchy residue theorem the left-hand side of (\ref{meixner3}) can be
rewritten as an integral over the contour $C_{1}$ (see Figure~1), namely,%
\begin{align}
&  \frac{1}{2\pi i}\int_{C_{1}}m_{n}^{\left(  \gamma,\ \mu\right)  }\left(
z\right)  \ m_{n}^{\left(  \gamma,\ \mu\right)  }\left(  z\right)
\ t^{z}\ \widetilde{\rho}\left(  z\right)  dz\label{int1}\\
&  \quad=\frac{\left(  1-t\right)  ^{x+y}}{\left(  1-\mu t\right)
^{x+y+\gamma}}\ \ _{2}F_{1}\left(
\begin{array}
[c]{c}%
-x,\ -y\\
\gamma
\end{array}
;\ \frac{\left(  1-\mu\right)  ^{2}t}{\mu\left(  1-t\right)  ^{2}}\right)
,\nonumber
\end{align}
where $\widetilde{\rho}\left(  z\right)  =\left(  \gamma\right)  _{z}%
\ \Gamma\left(  -z\right)  \left(  -\mu\right)  ^{z}$ and $\left\vert
t\right\vert \,<1.$ On the semicircle $z=-\gamma/2+R\ e^{i\theta}$ with
$-\pi/2\leq\theta\leq\pi/2$ the following estimate%
\begin{equation}
\widetilde{\rho}\left(  z\right)  =\text{O}\left(  R^{\gamma-1}\exp\left(
R\left(  \cos\theta\ln\left\vert \mu\right\vert -\sin\theta\left(  \arg\left(
-\mu\right)  \pm\pi\right)  \right)  \right)  \right)  \label{int2}%
\end{equation}
holds as $R\rightarrow\infty$ \cite{At:SusCH}. Therefore for $\left\vert
\mu\right\vert <1$ and $\left\vert \arg\left(  -\mu\right)  \right\vert <\pi$
the contour $C_{1}$ in (\ref{int1}) can be replaced by the contour $C_{2}$
where $z=-\gamma/2+ix$ and $-\infty<x<\infty.$ In view of the estimate
(\ref{int2}), when $\left\vert \arg\left(  -\mu\right)  \right\vert <\pi$ the
integral in (\ref{int1}) converges uniformly on the contour $C_{2},$ where
$\theta=\pm\pi/2,$ for all values of $\left\vert \mu\right\vert .$ As a
result, this integral can be analytically continued in the parameter $\mu$ to
the entire complex $\mu$-plane with the cut along the positive real axis
$\operatorname{Re}\mu>0.$ In particular, equation (\ref{int1}) remains valid
for both $\mu=\exp\left(  -2i\varphi\right)  $ and $z=-\gamma/2+ix$ $\left(
\gamma>0,0<\varphi<\pi\right)  .$ The substitution (\ref{pollaczek1}) results
in the integral (\ref{pollaczek6}) evaluation when $t\rightarrow e^{-2t}.$

\section{Solution of the Initial Value Problem}

We can now construct an explicit solution to the original Cauchy problem for
the time-dependent Schr\"{o}dinger equation of a modified oscillator in
(\ref{i7})--(\ref{i9}). More precisely, we will solve the following partial
differential equation%
\begin{align}
i\frac{\partial\psi}{\partial t} &  =\frac{1}{2}\sum_{s=1}^{n}\left(  -\left(
1+\cos2t\right)  \ \frac{\partial^{2}\psi}{\partial x_{s}^{2}}+\left(
1-\cos2t\right)  \ x_{s}^{2}\psi\right)  \label{sol0a}\\
&  \quad-\frac{i}{2}\sin2t\ \sum_{s=1}^{n}\left(  2x_{s}\frac{\partial\psi
}{\partial x_{s}}+\ \psi\right)  \nonumber
\end{align}
subject to the initial condition%
\begin{equation}
\left.  \psi\left(  \boldsymbol{x},t\right)  \right\vert _{t=0}=\psi
_{0}\left(  \boldsymbol{x}\right)  ,\qquad\boldsymbol{x}\in\boldsymbol{R}%
^{n}\label{sol0b}%
\end{equation}
by using the eigenfunction expansion (\ref{separ11})--(\ref{separ12}). Looking
for a particular solution of the system (\ref{separ7}) in the form%
\begin{equation}
u_{m}\left(  t\right)  =e^{-2i\xi t}\ p_{m}\left(  \xi\right)  ,\label{sol1}%
\end{equation}
where $\xi$ is a spectral parameter, one gets%
\begin{equation}
2\xi p_{m}\left(  \xi\right)  =\left(  m+j\right)  p_{m-1}\left(  \xi\right)
+\left(  m-j\right)  p_{m+1}\left(  \xi\right)  \label{sol2}%
\end{equation}
which coincides with the recurrence relation for the Meixner--Pollaczek
polynomials (\ref{pollaczek2}) when $\lambda=j+1$ and $\varphi=\pi/2.$ Thus
\begin{equation}
p_{m}\left(  \xi\right)  =P_{m-j-1}^{j+1}\left(  \xi,\ \frac{\pi}{2}\right)
.\label{sol3}%
\end{equation}
In view of the orthogonality relation (\ref{pollaczek3}), the Green function
$u_{mm^{\prime}}\left(  t\right)  ,$ or solution of the linear system
(\ref{separ7}) that satisfies the initial condition $u_{mm^{\prime}}\left(
0\right)  =\delta_{mm^{\prime}},$ can be obtain as the Fourier integral over
the spectral parameter%
\begin{align}
u_{mm^{\prime}}\left(  t\right)   &  =\frac{1}{d_{m^{\prime}}^{2}}%
\int_{-\infty}^{\infty}e^{-2i\xi t}\ p_{m}\left(  \xi\right)  p_{m^{\prime}%
}\left(  \xi\right)  \ \rho\left(  \xi\right)  d\xi\label{sol4}\\
&  =\frac{\left(  m^{\prime}-j-1\right)  !}{\left(  m^{\prime}+j\right)
!}\frac{2^{2j+2}}{2\pi}\nonumber\\
&  \quad\times\int_{-\infty}^{\infty}e^{-2i\xi t}\ P_{m-j-1}^{j+1}\left(
\xi,\frac{\pi}{2}\right)  P_{m^{\prime}-j-1}^{j+1}\left(  \xi,\frac{\pi}%
{2}\right)  \ \left\vert \Gamma\left(  j+1+i\xi\right)  \right\vert ^{2}%
d\xi\ \nonumber\\
&  =\frac{\left(  m+j\right)  !}{\left(  m-j-1\right)  !}\frac{\left(
-i\right)  ^{\left(  m-j-1\right)  +\left(  m^{\prime}-j-1\right)  }}%
{\Gamma\left(  2j+2\right)  }\frac{\left(  \sinh t\right)  ^{m+m^{\prime
}-2j-2}}{\left(  \cosh t\right)  ^{m+m^{\prime}}}\nonumber\\
&  \quad\times\ \ _{2}F_{1}\left(
\begin{array}
[c]{c}%
-m+j+1,\ -m^{\prime}+j+1\\
2j+2
\end{array}
;\ -\frac{1}{\sinh^{2}t}\right)  ,\nonumber
\end{align}
where the last integral has been evaluated with the help of the integral
representation (\ref{pollaczek6}).\smallskip

The generalized spherical harmonics for the the discrete positive series
$\mathcal{D}_{+}^{j}$ of the non-compact Lorentz group $SU\left(  1,1\right)
$ are \cite{Bargmann47}, \cite{Ni:Su:Uv}%
\begin{equation}
T_{mm^{\prime}}^{j}\left(  \alpha,\tau,\gamma\right)  =e^{-im\alpha
}v_{mm^{\prime}}^{j}\left(  \tau\right)  e^{-im^{\prime}\gamma}, \label{barg1}%
\end{equation}
where the Bargmann functions $v_{mm^{\prime}}^{j}\left(  \tau\right)  $ are
given by%
\begin{align}
&  v_{mm^{\prime}}^{j}\left(  \tau\right)  =\langle jm|e^{-i\tau J_{y}%
}|jm^{\prime}\rangle=e^{-n\tau/4}\int_{0}^{\infty}R_{NK}\left(  r\right)
R_{NK^{\prime}}\left(  e^{-\tau/2}r\right)  \ r^{n-1}dr\label{barg2}\\
&  \ \ \quad\qquad=\frac{\left(  -1\right)  ^{m-j-1}}{\Gamma\left(
2j+2\right)  }\sqrt{\frac{\left(  m+j\right)  !\left(  m^{\prime}+j\right)
!}{\left(  m-j-1\right)  !\left(  m^{\prime}-j-1\right)  !}}\ \left(
\sinh\frac{\tau}{2}\right)  ^{-2j-2}\left(  \tanh\frac{\tau}{2}\right)
^{m+m^{\prime}}\nonumber\\
&  \ \ \ \ \qquad\qquad\times\ \ _{2}F_{1}\left(
\begin{array}
[c]{c}%
-m+j+1,\ -m^{\prime}+j+1\\
2j+2
\end{array}
;\ -\frac{1}{\sinh^{2}\left(  \tau/2\right)  }\right)  .\nonumber
\end{align}
This implies the symmetry relation%
\begin{equation}
v_{mm^{\prime}}^{j}\left(  \tau\right)  =\left(  -1\right)  ^{m-m^{\prime}%
}v_{m^{\prime}m}^{j}\left(  \tau\right)  \label{barg3}%
\end{equation}
and the differentiation formula%
\begin{align}
&  2\frac{d}{d\tau}v_{m^{\prime}m}^{j}\left(  \tau\right)  =\sqrt{\left(
m-j-1\right)  \left(  m+j\right)  }\ v_{m^{\prime},m-1}^{j}\left(  \tau\right)
\label{barg4}\\
&  \quad\qquad\qquad\quad-\sqrt{\left(  m+j+1\right)  \left(  m-j\right)
}\ v_{m^{\prime},m+1}^{j}\left(  \tau\right)  ,\nonumber
\end{align}
which follows directly from (\ref{barg2}) and (\ref{harm14}). Also%
\begin{equation}
\sum_{m^{\prime\prime}=j+1}^{\infty}v_{mm^{\prime\prime}}^{j}\left(
\tau\right)  v_{m^{\prime}m^{\prime\prime}}^{j}\left(  \tau\right)
=\delta_{mm^{\prime}}, \label{barg5}%
\end{equation}
see \cite{Bargmann47}, \cite{Ni:Su:Uv}, and \cite{Smir:Sus:Shir} for more
details.\smallskip

Combining equations (\ref{separ11})--(\ref{separ12}), (\ref{sol4}) and
(\ref{barg2})--(\ref{barg3}) together, we finally\ arrive at the following
expression in terms of the Bargmann functions%
\begin{equation}
c_{m}\left(  t\right)  =e^{-i\omega mt}\sum_{m^{\prime}=j+1}^{\infty
}i^{m^{\prime}-m}v_{m^{\prime}m}^{j}\left(  2t\right)  \ \int_{\boldsymbol{R}%
^{n}}\psi_{jm^{\prime}\left\{  \nu\right\}  }^{\ast}\left(  \boldsymbol{x}%
^{\prime}\right)  \psi_{0}\left(  \boldsymbol{x}^{\prime}\right)
\ dv^{\prime}\label{sol5}%
\end{equation}
for the time-depending coefficients $c_{m}\left(  t\right)  $ in the expansion
(\ref{separ11}), namely,%
\begin{equation}
\psi\left(  \boldsymbol{x},t\right)  =\sum_{j\left\{  \nu\right\}  }%
\sum_{m=j+1}^{\infty}c_{m}\left(  t\right)  \ \psi_{jm\left\{  \nu\right\}
}\left(  \boldsymbol{x}\right)  ,\label{sol6}%
\end{equation}
for the solution of the original initial value problem (\ref{i7})--(\ref{i9})
with $\omega=2.$ With the help of the differentiation formula (\ref{barg4})
one can easily verify that these coefficients $c_{m}\left(  t\right)  $
satisfy the system of ordinary differential equations (\ref{separ3}) with
corresponding initial conditions. This gives a direct proof that our solution
$\psi\left(  \boldsymbol{x},t\right)  $ does satisfy the initial value problem
(\ref{i7})--(\ref{i9}).\smallskip

Thus constructed solution $\psi\left(  \boldsymbol{x},t\right)  $ belongs to
the space of the square integrable functions $\mathcal{L}^{2}\left(
\boldsymbol{R}^{n}\right)  $ for all times provided that the initial function
$\psi_{0}\left(  \boldsymbol{x}\right)  $ is of the same class $\psi_{0}%
\in\mathcal{L}^{2}\left(  \boldsymbol{R}^{n}\right)  .$ Indeed, in this case
the condition%
\begin{equation}
\sum_{m=j+1}^{\infty}\left\vert c_{m}\left(  t\right)  \right\vert ^{2}=1
\label{sol7}%
\end{equation}
holds for all values of $t$ in view of the unitary relation (\ref{barg5}) of
Bargmann's functions. We shall take advantage of the group-theoretical meaning
of this solution in order to cunstruct the corresponding Green function later.

\section{The Finite ``Rotation'' Operators}

We apply a general approach to a certain type of the integral transforms
\cite{As:Ra:Su}, \cite{Ra:Su}, \cite{SusAlg}, \cite{SuslovBFS}, which was
originated by Wiener \cite{Wi}, to our problem. Consider the following
bilinear sum%
\begin{align}
S_{t}\left(  r,r^{\prime}\right)   &  =\sum_{N=K,K+2,...\ ,\infty}%
R_{NK}\left(  r\right)  R_{NK}\left(  r^{\prime}\right)  \ t^{\left(
N-K\right)  /2}\label{bilrosc1}\\
&  =2\exp\left(  -\left(  r^{2}+r^{\prime2}\right)  /2\right)  \ \left(
rr^{\prime}\right)  ^{K}\nonumber\\
&  \quad\times\sum_{k=0}^{\infty}\frac{k!}{\Gamma\left(  K+n/2+k\right)
}\ L_{k}^{K+n/2-1}\left(  r^{2}\right)  L_{k}^{K+n/2-1}\left(  \left.
r^{\prime}\right.  ^{2}\right)  \ t^{k}\nonumber
\end{align}
for the oscillator radial functions (\ref{harm7}). With the help of the
Poisson kernel for the Laguerre polynomials%
\begin{align}
&  \sum_{k=0}^{\infty}\frac{k!}{\left(  \alpha+1\right)  _{k}}\ L_{k}^{\alpha
}\left(  x\right)  L_{k}^{\alpha}\left(  y\right)  \ t^{k}%
\label{LaguerrePoisson}\\
&  \quad=\left(  1-t\right)  ^{-\alpha-1}\exp\left(  -\frac{\left(
x+y\right)  t}{1-t}\right)  \ \ _{0}F_{1}\left(
\begin{array}
[c]{c}%
-\\
\alpha+1
\end{array}
;\ \frac{xyt}{\left(  1-t\right)  ^{2}}\right)  ,\nonumber
\end{align}
see, for example, \cite{Rainville}, p.~212; we derive the following closed
form%
\begin{align}
&  S_{t}\left(  r,r^{\prime}\right)  =\frac{2}{\Gamma\left(  K+n/2\right)
}\left(  rr^{\prime}\right)  ^{K}\left(  1-t\right)  ^{-\left(  K+n/2\right)
}\exp\left(  -\frac{r^{2}+\left.  r^{\prime}\right.  ^{2}}{2}\frac{1+t}%
{1-t}\right) \nonumber\\
&  \quad\quad\qquad\quad\times\ \ _{0}F_{1}\left(
\begin{array}
[c]{c}%
-\\
K+n/2
\end{array}
;\ \frac{\left(  rr^{\prime}\right)  ^{2}t}{\left(  1-t\right)  ^{2}}\right)
,\qquad\left\vert t\right\vert \leq1,\quad t\neq1 \label{bilrosc2}%
\end{align}
for this kernel. In the case $t=e^{i\alpha}$ one gets%
\begin{align}
&  S_{e^{i\alpha}}\left(  r,r^{\prime}\right)  =\frac{2^{-2j-1}e^{i\left(
\pi-\alpha\right)  \left(  j+1\right)  }}{\Gamma\left(  2j+2\right)  \left(
\sin\left(  \alpha/2\right)  \right)  ^{2j+2}}\left(  rr^{\prime}\right)
^{2j+2-n/2}\exp\left(  \frac{r^{2}+\left.  r^{\prime}\right.  ^{2}}%
{2i\tan\left(  \alpha/2\right)  }\right) \nonumber\\
&  \quad\quad\qquad\quad\quad\times\ \ _{0}F_{1}\left(
\begin{array}
[c]{c}%
-\\
2j+2
\end{array}
;\ -\frac{\left(  rr^{\prime}\right)  ^{2}}{4\sin^{2}\left(  \alpha/2\right)
}\right)  , \label{bilrosc3}%
\end{align}
where we use the $SU\left(  1,1\right)  $ moment $j=K/2+n/4-1.$\smallskip

Using the orthogonality property of the radial functions%
\begin{equation}
\int_{0}^{\infty}R_{NK}\left(  r\right)  R_{N^{\prime}K}\left(  r\right)
\ r^{n-1}dr=\delta_{NN^{\prime}}, \label{bilrosc4}%
\end{equation}
from (\ref{bilrosc1}) we obtain the following integral equation%
\begin{equation}
t^{\left(  N-K\right)  /2}R_{NK}\left(  r\right)  =\int_{0}^{\infty}%
S_{t}\left(  r,r^{\prime}\right)  R_{NK}\left(  r^{\prime}\right)  \ \left(
r^{\prime}\right)  ^{n-1}dr^{\prime},\quad\left\vert t\right\vert <1.
\label{bilrosc5}%
\end{equation}
In the $SU\left(  1,1\right)  $ notations for the oscillator wave
functions(\ref{harm12}), when $t=e^{i\alpha},$ it takes the form%
\begin{equation}
e^{im\alpha}\psi_{jm\left\{  \nu\right\}  }\left(  \boldsymbol{x}\right)
=\int_{0}^{\infty}G_{\alpha}^{j}\left(  r,r^{\prime}\right)  \psi_{jm\left\{
\nu\right\}  }\left(  \boldsymbol{x}^{\prime}\right)  \ \left(  r^{\prime
}\right)  ^{n-1}dr^{\prime}, \label{bilrosc6}%
\end{equation}
where%
\begin{align}
G_{\alpha}^{j}\left(  r,r^{\prime}\right)   &  =\frac{2^{-2j-1}e^{i\pi\left(
j+1\right)  }}{\Gamma\left(  2j+2\right)  \left(  \sin\left(  \alpha/2\right)
\right)  ^{2j+2}}\left(  rr^{\prime}\right)  ^{2j+2-n/2}\exp\left(
\frac{r^{2}+\left.  r^{\prime}\right.  ^{2}}{2i\tan\left(  \alpha/2\right)
}\right) \nonumber\\
&  \quad\qquad\times\ \ _{0}F_{1}\left(
\begin{array}
[c]{c}%
-\\
2j+2
\end{array}
;\ -\frac{\left(  rr^{\prime}\right)  ^{2}}{4\sin^{2}\left(  \alpha/2\right)
}\right)  \label{bilrosc6a}%
\end{align}
and $\boldsymbol{x}/r=\boldsymbol{x}^{\prime}/r^{\prime}=\boldsymbol{n}%
=\boldsymbol{n}\left(  \Omega\right)  $ with $\boldsymbol{n}^{2}=1.$ The
following symmetry properties hold%
\begin{equation}
G_{\alpha}^{j}\left(  r,r^{\prime}\right)  =G_{\alpha}^{j}\left(  r^{\prime
},r\right)  =\left(  G_{-\alpha}^{j}\left(  r,r^{\prime}\right)  \right)
^{\ast} \label{bilrosc6b}%
\end{equation}
and the formal orthogonality relation is%
\begin{equation}
\int_{r^{\prime\prime}=0}^{\infty}\left(  G_{\alpha}^{j}\left(  r,r^{\prime
\prime}\right)  \right)  ^{\ast}G_{\alpha}^{j}\left(  r^{\prime\prime
},r^{\prime}\right)  \ \left(  r^{\prime\prime}\right)  ^{n-1}dr^{\prime
\prime}=\frac{\delta\left(  r-r^{\prime}\right)  }{r^{n-1}}, \label{bilrosc6c}%
\end{equation}
where $\delta\left(  r\right)  $ is the Dirac delta function. Also,%
\begin{equation}
\sum_{m=j+1}^{\infty}e^{im\alpha}\psi_{jm\left\{  \nu\right\}  }\left(
\boldsymbol{x}\right)  \psi_{jm\left\{  \nu\right\}  }^{\ast}\left(
\boldsymbol{x}^{\prime}\right)  =Y_{K\nu}\left(  \Omega\right)  \ Y_{K\nu
}\left(  \Omega^{\prime}\right)  \ G_{\alpha}^{j}\left(  r,r^{\prime}\right)
,\qquad\alpha\neq0. \label{bilrosc6d}%
\end{equation}
In particular, when $\alpha=\pm\pi/2,$ one gets%
\begin{equation}
\left(  \pm i\right)  ^{m}\psi_{jm\left\{  \nu\right\}  }\left(
\boldsymbol{x}\right)  =\int_{0}^{\infty}G_{\pm\pi/2}^{j}\left(  r,r^{\prime
}\right)  \psi_{jm\left\{  \nu\right\}  }\left(  \boldsymbol{x}^{\prime
}\right)  \ \left(  r^{\prime}\right)  ^{n-1}dr^{\prime} \label{bilrosc7}%
\end{equation}
with%
\begin{equation}
G_{\pm\pi/2}^{j}\left(  r,r^{\prime}\right)  =\frac{e^{\pm i\pi\left(
j+1\right)  }\left(  rr^{\prime}\right)  ^{2j+2-n/2}}{2^{j}\Gamma\left(
2j+2\right)  }\ e^{\pm\left(  r^{2}+\left.  r^{\prime}\right.  ^{2}\right)
/2i}\ _{0}F_{1}\left(
\begin{array}
[c]{c}%
-\\
2j+2
\end{array}
;\ -\frac{1}{2}\left(  rr^{\prime}\right)  ^{2}\right)  \label{bilrosc8}%
\end{equation}
These formulas will allow us to find a different form of the solution
(\ref{sol5})--(\ref{sol6}) in the next section.\smallskip

In the process we found out that the finite \textquotedblleft
rotation\textquotedblright\ operator $e^{-i\alpha J_{z}}$ of the group
$SU\left(  1,1\right)  $ acts on the oscillator wave functions (\ref{harm12})
as the following integral operator%
\begin{align}
e^{-i\alpha J_{z}}\psi_{jm\left\{  \nu\right\}  }\left(  \boldsymbol{x}%
\right)   &  =\int_{0}^{\infty}G_{-\alpha}^{j}\left(  r,r^{\prime}\right)
\psi_{jm\left\{  \nu\right\}  }\left(  \boldsymbol{x}^{\prime}\right)
\ \left(  r^{\prime}\right)  ^{n-1}dr^{\prime}\label{bilrosc9}\\
&  =Y_{K\nu}\left(  \Omega\right)  \int_{0}^{\infty}G_{-\alpha}^{j}\left(
r,r^{\prime}\right)  R_{NK}\left(  r^{\prime}\right)  \ \left(  r^{\prime
}\right)  ^{n-1}dr^{\prime}\nonumber
\end{align}
with the kernel explicitly given by (\ref{bilrosc6a}). Also, in view of
(\ref{barg2}),%
\begin{align}
e^{-i\tau J_{y}}\psi_{jm\left\{  \nu\right\}  }\left(  \boldsymbol{x}\right)
&  =\sum_{m^{\prime}=j+1}^{\infty}v_{m^{\prime}m}^{j}\left(  \tau\right)
\ \psi_{jm^{\prime}\left\{  \nu\right\}  }\left(  \boldsymbol{x}\right)
\label{bilrosc10}\\
&  =e^{-n\tau/4}\psi_{jm\left\{  \nu\right\}  }\left(  e^{-\tau/2}%
\boldsymbol{x}\right)  =e^{-n\tau/4}Y_{K\nu}\left(  \Omega\right)
R_{NK}\left(  e^{-\tau/2}r\right)  ,\nonumber
\end{align}
see \cite{Ni:Su:Uv} for more details.

\section{An Integral Form of the Solution}

We rewrite the coefficients (\ref{sol5}) in the form%
\begin{equation}
c_{m}\left(  t\right)  =e^{-im\left(  \omega t+\pi/2\right)  }\sum_{m^{\prime
}=j+1}^{\infty}v_{m^{\prime}m}^{j}\left(  2t\right)  \ \int_{\boldsymbol{R}%
^{n}}\left(  \left(  -i\right)  ^{m^{\prime}}\psi_{jm^{\prime}\left\{
\nu\right\}  }\left(  \boldsymbol{x}^{\prime}\right)  \right)  ^{\ast}\psi
_{0}\left(  \boldsymbol{x}^{\prime}\right)  \ dv^{\prime}\label{gr1}%
\end{equation}
and use the integral transform (\ref{bilrosc7}) as
\begin{equation}
\left(  -i\right)  ^{m^{\prime}}\psi_{jm^{\prime}\left\{  \nu\right\}
}\left(  \boldsymbol{x}^{\prime}\right)  =\int_{0}^{\infty}G_{-\pi/2}%
^{j}\left(  r^{\prime},r^{\prime\prime}\right)  \ \psi_{jm^{\prime}\left\{
\nu\right\}  }\left(  \boldsymbol{x}^{\prime\prime}\right)  \ \left(
r^{\prime\prime}\right)  ^{n-1}dr^{\prime\prime},\label{gr2}%
\end{equation}
where $\boldsymbol{x}^{\prime}/r^{\prime}=\boldsymbol{x}^{\prime\prime
}/r^{\prime\prime}=\boldsymbol{n}^{\prime}=\boldsymbol{n}\left(
\Omega^{\prime}\right)  ,$ in order to obtain%
\begin{align}
c_{m}\left(  t\right)   &  =e^{-im\left(  \omega t+\pi/2\right)
}\ \label{gr3}\\
&  \times\int_{\boldsymbol{R}^{n}}\left(  \int_{0}^{\infty}G_{-\pi/2}%
^{j}\left(  r^{\prime},r^{\prime\prime}\right)  \left(  \sum_{m^{\prime}%
=j+1}^{\infty}v_{m^{\prime}m}^{j}\left(  2t\right)  \ \psi_{jm^{\prime
}\left\{  \nu\right\}  }\left(  \boldsymbol{x}^{\prime\prime}\right)  \right)
\ \left(  r^{\prime\prime}\right)  ^{n-1}dr^{\prime\prime}\right)  ^{\ast}%
\psi_{0}\left(  \boldsymbol{x}^{\prime}\right)  \ dv^{\prime}\nonumber\\
&  =e^{-im\left(  \omega t+\pi/2\right)  }\int_{\boldsymbol{R}^{n}}\left(
\int_{0}^{\infty}G_{-\pi/2}^{j}\left(  r^{\prime},r^{\prime\prime}\right)
\left(  e^{-2itJ_{y}}\ \psi_{jm\left\{  \nu\right\}  }\left(  \boldsymbol{x}%
^{\prime\prime}\right)  \ \right)  \left(  r^{\prime\prime}\right)
^{n-1}dr^{\prime\prime}\right)  ^{\ast}\psi_{0}\left(  \boldsymbol{x}^{\prime
}\right)  \ dv^{\prime}\nonumber\\
&  =e^{-im\left(  \omega t+\pi/2\right)  -nt/2}\int_{\boldsymbol{R}^{n}%
}\left(  \int_{0}^{\infty}G_{\pi/2}^{j}\left(  r^{\prime},r^{\prime\prime
}\right)  \ \psi_{jm\left\{  \nu\right\}  }^{\ast}\left(  e^{-t}%
\boldsymbol{x}^{\prime\prime}\right)  \ \left(  r^{\prime\prime}\right)
^{n-1}dr^{\prime\prime}\right)  \ \psi_{0}\left(  \boldsymbol{x}^{\prime
}\right)  \ dv^{\prime}\nonumber
\end{align}
with the help of (\ref{bilrosc10}). Substitution into the eigenfunction
expansion (\ref{sol6}) gives%
\begin{align}
&  \psi\left(  \boldsymbol{x},t\right)  =e^{-nt/2}\int_{\boldsymbol{R}^{n}%
}\ \psi_{0}\left(  \boldsymbol{x}^{\prime}\right)  \label{gr4}\\
&  \times\ \left(  \sum_{j\left\{  \nu\right\}  }\int_{0}^{\infty}G_{\pi
/2}^{j}\left(  r^{\prime},r^{\prime\prime}\right)  \ \left(  \sum
_{m=j+1}^{\infty}e^{-im\left(  \omega t+\pi/2\right)  }\psi_{jm\left\{
\nu\right\}  }\left(  \boldsymbol{x}\right)  \psi_{jm\left\{  \nu\right\}
}^{\ast}\left(  e^{-t}\boldsymbol{x}^{\prime\prime}\right)  \right)  \ \left(
r^{\prime\prime}\right)  ^{n-1}dr^{\prime\prime}\right)  \ dv^{\prime
},\nonumber
\end{align}
where by (\ref{harm12}) and (\ref{bilrosc6d})%
\begin{equation}
\sum_{m=j+1}^{\infty}e^{-im\left(  \omega t+\pi/2\right)  }\psi_{jm\left\{
\nu\right\}  }\left(  \boldsymbol{x}\right)  \psi_{jm\left\{  \nu\right\}
}^{\ast}\left(  e^{-t}\boldsymbol{x}^{\prime\prime}\right)  =Y_{K\nu}\left(
\Omega\right)  \ Y_{K\nu}^{\ast}\left(  \Omega^{\prime}\right)  \ G_{-\omega
t-\pi/2}^{j}\left(  r,e^{-t}r^{\prime\prime}\right)  ,\label{gr5}%
\end{equation}
and our solution takes the form%
\begin{align}
&  \psi\left(  \boldsymbol{x},t\right)  =e^{-nt/2}\int_{\boldsymbol{R}^{n}%
}\ \psi_{0}\left(  \boldsymbol{x}^{\prime}\right)  \label{gr6}\\
&  \times\ \left(  \sum_{j\left\{  \nu\right\}  }Y_{K\nu}\left(
\Omega\right)  \ Y_{K\nu}^{\ast}\left(  \Omega^{\prime}\right)  \ \int
_{0}^{\infty}G_{\pi/2}^{j}\left(  r^{\prime},r^{\prime\prime}\right)
\ G_{-\omega t-\pi/2}^{j}\left(  r,e^{-t}r^{\prime\prime}\right)  \ \left(
r^{\prime\prime}\right)  ^{n-1}dr^{\prime\prime}\right)  \ dv^{\prime
}\nonumber
\end{align}
with $\omega=2.$ The last integral can be evaluated with the help of the
formula (\ref{iap5}) from the appendix at the end of the paper%
\begin{align}
&  \int_{0}^{\infty}G_{\pi/2}^{j}\left(  r^{\prime},r^{\prime\prime}\right)
\ \left(  G_{2t+\pi/2}^{j}\left(  r,e^{-t}r^{\prime\prime}\right)  \right)
^{\ast}\ \left(  r^{\prime\prime}\right)  ^{n-1}dr^{\prime\prime}\label{gr7}\\
&  \quad=e^{nt/2}\frac{e^{i\pi\left(  j+1\right)  }}{2^{j}\Gamma\left(
2j+1\right)  }\frac{\left(  rr^{\prime}\right)  ^{2j+2-n/2}}{\left(
e^{-t}\cos\left(  t+\pi/4\right)  -e^{t}\sin\left(  t+\pi/4\right)  \right)
^{2j+2}}\nonumber\\
&  \qquad\times\exp\left(  \frac{r^{2}}{2i}\frac{e^{t}\cos\left(
t+\pi/4\right)  +e^{-t}\sin\left(  t+\pi/4\right)  }{e^{-t}\cos\left(
t+\pi/4\right)  -e^{t}\sin\left(  t+\pi/4\right)  }\right)  \nonumber\\
&  \qquad\times\exp\left(  \frac{\left(  r^{\prime}\right)  ^{2}}{2i}%
\frac{e^{-t}\cos\left(  t+\pi/4\right)  +e^{t}\sin\left(  t+\pi/4\right)
}{e^{-t}\cos\left(  t+\pi/4\right)  -e^{t}\sin\left(  t+\pi/4\right)
}\right)  \nonumber\\
&  \qquad\times~_{0}F_{1}\left(
\begin{array}
[c]{c}%
-\\
2j+2
\end{array}
;\ -\frac{\left(  rr^{\prime}\right)  ^{2}}{2\left(  e^{-t}\cos\left(
t+\pi/4\right)  -e^{t}\sin\left(  t+\pi/4\right)  \right)  ^{2}}\right)
.\nonumber
\end{align}

As a result, the solution of the Cauchy initial value problem (\ref{i7}%
)--(\ref{i9}) is given by%
\begin{equation}
\psi\left(  \boldsymbol{x},t\right)  =\int_{\boldsymbol{R}^{n}}G_{t}\left(
\boldsymbol{x},\boldsymbol{x}^{\prime}\right)  \ \psi_{0}\left(
\boldsymbol{x}^{\prime}\right)  \ dv^{\prime}, \label{gr8}%
\end{equation}
where the Green function is%
\begin{equation}
G_{t}\left(  \boldsymbol{x},\boldsymbol{x}^{\prime}\right)  =\sum_{K\nu
}Y_{K\nu}\left(  \Omega\right)  \ Y_{K\nu}^{\ast}\left(  \Omega^{\prime
}\right)  \ \mathcal{G}_{t}^{K}\left(  r,r^{\prime}\right)  \label{gr9}%
\end{equation}
with%
\begin{align}
\mathcal{G}_{t}^{K}\left(  r,r^{\prime}\right)   &  =\frac{e^{-i\pi\left(
2K+n\right)  /4}}{2^{K+n/2-1}\Gamma\left(  K+n/2\right)  }\ \frac{\left(
rr^{\prime}\right)  ^{K}}{\left(  \cos t\sinh t+\sin t\cosh t\right)
^{K+n/2}}\label{gr10}\\
&  \quad\times\exp\left(  i\frac{\left(  r^{2}+\left(  r^{\prime}\right)
^{2}\right)  \cos t\cosh t-\left(  r^{2}-\left(  r^{\prime}\right)
^{2}\right)  \sin t\sinh t}{2\left(  \cos t\sinh t+\sin t\cosh t\right)
}\right) \nonumber\\
&  \quad\times~_{0}F_{1}\left(
\begin{array}
[c]{c}%
-\\
K+n/2
\end{array}
;\ -\frac{\left(  rr^{\prime}\right)  ^{2}}{4\left(  \cos t\sinh t+\sin t\cosh
t\right)  ^{2}}\right)  .\nonumber
\end{align}
The details of the calculations are left to the reader.\smallskip

The Green function can also be independently found by separation of the
variables in the Cartesian coordinates. Indeed, when $n=1$ and $K=0,1$ with
the help of the familiar relations%
\begin{equation}
\cos\alpha=~_{0}F_{1}\left(
\begin{array}
[c]{c}%
-\\
1/2
\end{array}
;\ -\frac{\alpha^{2}}{4}\right)  ,\qquad\sin\alpha=\alpha\ _{0}F_{1}\left(
\begin{array}
[c]{c}%
-\\
3/2
\end{array}
;\ -\frac{\alpha^{2}}{4}\right)  \label{gr11}%
\end{equation}
our equations (\ref{gr9})--(\ref{gr10}) can be reduced to%
\begin{align}
&  G_{t}\left(  x,x^{\prime}\right)  =\frac{1}{2}\left(  \mathcal{G}_{t}%
^{0}\left(  x,x^{\prime}\right)  +\mathcal{G}_{t}^{1}\left(  x,x^{\prime
}\right)  \right) \label{gr12}\\
&  \quad\quad\quad\quad=\frac{1}{\sqrt{2\pi i\left(  \cos t\sinh t+\sin t\cosh
t\right)  }}\nonumber\\
&  \qquad\quad\qquad\times\exp\left(  \frac{\left(  x^{2}-\left(  x^{\prime
}\right)  ^{2}\right)  \sin t\sinh t+2xx^{\prime}-\left(  x^{2}+\left(
x^{\prime}\right)  ^{2}\right)  \cos t\cosh t}{2i\left(  \cos t\sinh t+\sin
t\cosh t\right)  }\right)  ,\nonumber
\end{align}
which gives the Green function for the one-dimensional Schr\"{o}dinger
equation (\ref{sol0a}); we shall elaborate on the one-dimensional case in the
next section. Thus, in the general case,%
\begin{align}
&  G_{t}\left(  \boldsymbol{x},\boldsymbol{x}^{\prime}\right)  =%
%TCIMACRO{\dprod _{s=1}^{n}}%
%BeginExpansion
{\displaystyle\prod_{s=1}^{n}}
%EndExpansion
G_{t}\left(  x_{s},x_{s}^{\prime}\right) \label{gr13}\\
&  \qquad\ \ \quad\ \ =\left(  \frac{1}{2\pi i\left(  \cos t\sinh t+\sin
t\cosh t\right)  }\right)  ^{n/2}\nonumber\\
\  &  \qquad\qquad\ \ \ \ \times\exp\left(  \frac{\left(  \boldsymbol{x}%
^{2}-\boldsymbol{x}^{\prime2}\right)  \sin t\sinh t+2\boldsymbol{x}%
\cdot\boldsymbol{x}^{\prime}-\left(  \boldsymbol{x}^{2}+\boldsymbol{x}%
^{\prime2}\right)  \cos t\cosh t}{2i\left(  \cos t\sinh t+\sin t\cosh
t\right)  }\right)  ,\nonumber
\end{align}
and equation (\ref{gr9}) gives an expansion formula for this Green function in
terms of the corresponding hyperspherical harmonics. This type of oscillatory
integrals is discussed in \cite{SteinHarm}.\smallskip

The time evolution operator for the time-dependent Schr\"{o}dinger equation
(\ref{i7}) can formally be written as%
\begin{equation}
U\left(  t,t_{0}\right)  =\mathbf{T}\left(  \exp\left(  -\frac{i}{\hslash}%
\int_{t_{0}}^{t}H\left(  t^{\prime}\right)  \ dt^{\prime}\right)  \right)  ,
\label{gr14}%
\end{equation}
where $\mathbf{T}$ is the time ordering operator which orders operators with
larger times to the left \cite{Bo:Shi}, \cite{Flu}. Namely, this unitary
operator takes a state at time $t_{0}$ to a state at time $t,$ so that%
\begin{equation}
\psi\left(  x,t\right)  =U\left(  t,t_{0}\right)  \psi\left(  x,t_{0}\right)
\label{gr15}%
\end{equation}
and%
\begin{equation}
U\left(  t,t_{0}\right)  =U\left(  t,t^{\prime}\right)  U\left(  t^{\prime
},t_{0}\right)  , \label{gr16}%
\end{equation}%
\begin{equation}
U^{-1}\left(  t,t_{0}\right)  =U^{\dagger}\left(  t,t_{0}\right)  =U\left(
t_{0},t\right)  . \label{gr17}%
\end{equation}
We have constructed this time evolution operator explicitly in (\ref{gr8}), as
the integral operator with the kernel given by (\ref{gr13}), for the
particular form of the time-dependent Hamiltonian operator of a modified
oscillator in (\ref{i8}).

\section{The Forced Modified Oscillator}

In the previous section, among other things, we have solved the following
one-dimensional time-depending Schr\"{o}dinger equation for a modified
oscillator%
\begin{equation}
i\frac{\partial\psi}{\partial t}=-\frac{1}{2}\left(  1+\cos2t\right)
\ \frac{\partial^{2}\psi}{\partial x^{2}}+\frac{1}{2}\left(  1-\cos2t\right)
\ x^{2}\psi-\frac{i}{2}\sin2t\ \left(  2x\frac{\partial\psi}{\partial
x}+\ \psi\right)  \label{fmo1}%
\end{equation}
on the infinite interval subject to the initial condition%
\begin{equation}
\left.  \psi\left(  x,t\right)  \right\vert _{t=0}=\psi_{0}\left(  x\right)
\qquad\left(  -\infty<x<\infty\right)  . \label{fmo2}%
\end{equation}
Our solution takes the form%
\begin{equation}
\psi\left(  x,t\right)  =\int_{-\infty}^{\infty}G_{0}\left(  x,y,t\right)
\ \psi_{0}\left(  y\right)  \ dy, \label{fmo3}%
\end{equation}
where the Green function (or Feynman's propagator) is given by%
\begin{align}
G_{0}\left(  x,y,t\right)   &  =\frac{1}{\sqrt{2\pi i\left(  \cos t\sinh
t+\sin t\cosh t\right)  }}\label{fmo4}\\
&  \quad\times\exp\left(  \frac{\left(  x^{2}-y^{2}\right)  \sin t\sinh
t+2xy-\left(  x^{2}+y^{2}\right)  \cos t\cosh t}{2i\left(  \cos t\sinh t+\sin
t\cosh t\right)  }\right)  .\nonumber
\end{align}
This expression may be considered as a generalization of the propagator for
the simple harmonic oscillator; see \cite{FeynmanPhD}, \cite{Feynman},
\cite{Fey:Hib}, \cite{Gottf:T-MY}, \cite{Holstein}, \cite{Merz}, and
references therein.\smallskip

In this section, we shall extend this solution to a more general case of the
forced modified oscillator with the Schr\"{o}dinger equation of the form%
\begin{align}
i\frac{\partial\psi}{\partial t} &  =-\frac{1}{2}\left(  1+\cos2t\right)
\ \frac{\partial^{2}\psi}{\partial x^{2}}+\frac{1}{2}\left(  1-\cos2t\right)
\ x^{2}\psi\label{fmo5}\\
&  \quad-\frac{i}{2}\sin2t\ \left(  2x\frac{\partial\psi}{\partial x}%
+\ \psi\right)  -f\left(  t\right)  \ x\psi+ig\left(  t\right)  \ \frac
{\partial\psi}{\partial x},\nonumber
\end{align}
where $f\left(  t\right)  $ and $g\left(  t\right)  $ are two arbitrary real
valued functions of time only. Indeed, by a method similar to one in
\cite{Lop:Sus} and \cite{Merz} for the case of the forced harmonic oscillator,
one can look for the Green function in the form%
\begin{equation}
\psi=u\ e^{iS},\label{fmo6}%
\end{equation}
where $u=G_{0}\left(  x,y,t\right)  $ is the fundamental solution of the
Schr\"{o}dinger equation for the modified oscillator (\ref{fmo1}) and
$S=\alpha\left(  t\right)  x+\beta\left(  t\right)  y+\gamma\left(  t\right)
.$ Its substitution into (\ref{fmo5}) results in%
\begin{align}
&  \left(  \frac{d\alpha}{dt}x+\frac{d\beta}{dt}y+\frac{d\gamma}{dt}\right)
u\label{fmo7}\\
&  \quad=\frac{1}{2}\left(  1+\cos2t\right)  \left(  -\alpha^{2}%
u+2i\alpha\frac{\partial u}{\partial x}\right)  -\sin2t\ \alpha
xu+fxu+g\left(  \alpha u-i\frac{\partial u}{\partial x}\right)  ,\nonumber
\end{align}
where by (\ref{fmo4})%
\begin{equation}
\frac{\partial u}{\partial x}=i\frac{x\left(  \cos t\cosh t-\sin t\sinh
t\right)  -y}{\cos t\sinh t+\sin t\cosh t}\ u.\label{fmo8}%
\end{equation}
Thus%
\begin{align}
&  \frac{d\alpha}{dt}x+\frac{d\beta}{dt}y+\frac{d\gamma}{dt}=-\sin
2t\ \alpha\ x+f\ x\label{fmo8a}\\
&  \quad-\frac{1}{2}\left(  1+\cos2t\right)  \left(  \alpha^{2}+2\alpha
\frac{x\left(  \cos t\cosh t-\sin t\sinh t\right)  -y}{\cos t\sinh t+\sin
t\cosh t}\right)  \nonumber\\
&  \qquad+g\left(  \alpha+\frac{x\left(  \cos t\cosh t-\sin t\sinh t\right)
-y}{\cos t\sinh t+\sin t\cosh t}\right)  \nonumber
\end{align}
and equating the coefficients of $x,$ $y$ and $1,$ we obtain the following
system of the first order differential equations%
\begin{equation}
\frac{d\alpha\left(  t\right)  }{dt}+\frac{2}{\tan t+\tanh t}\ \alpha\left(
t\right)  =f\left(  t\right)  +g\left(  t\right)  \frac{1-\tan t\ \tanh
t}{\tan t+\tanh t},\label{fmo9}%
\end{equation}%
\begin{equation}
\frac{d\beta\left(  t\right)  }{dt}=\frac{\left(  1+\cos2t\right)
\alpha\left(  t\right)  -g\left(  t\right)  }{\cos t\sinh t+\sin t\cosh
t},\label{fmo10}%
\end{equation}
and%
\begin{equation}
\frac{d\gamma\left(  t\right)  }{dt}=\alpha\left(  t\right)  g\left(
t\right)  -\frac{1}{2}\left(  1+\cos2t\right)  \alpha^{2}\left(  t\right)
.\label{fmo11}%
\end{equation}
In view of%
\begin{equation}
\frac{\mu^{\prime}\left(  t\right)  }{\mu\left(  t\right)  }=\frac{\ 2}{\tan
t+\tanh t},\quad\mu\left(  t\right)  =\cos t\sinh t+\sin t\cosh
t,\label{fmo11a}%
\end{equation}
equation (\ref{fmo9}) takes the form%
\begin{equation}
\frac{d}{dt}\left(  \mu\left(  t\right)  \alpha\left(  t\right)  \right)
=\mu\left(  t\right)  \left(  f\left(  t\right)  +g\left(  t\right)
\frac{1-\tan t\ \tanh t}{\tan t+\tanh t}\right)  .\label{fmo12}%
\end{equation}
The solutions are%
\begin{align}
\alpha\left(  t\right)   &  =\left(  \cos t\sinh t+\sin t\cosh t\right)
^{-1}\label{fmo13}\\
&  \times\int_{0}^{t}\left(  f\left(  s\right)  \left(  \cos s\sinh s+\sin
s\cosh s\right)  +g\left(  s\right)  \left(  \cos s\cosh s-\sin s\sinh
s\right)  \right)  \ ds,\nonumber
\end{align}%
\begin{equation}
\beta\left(  t\right)  =\int_{0}^{t}\frac{\left(  1+\cos2s\right)
\alpha\left(  s\right)  -g\left(  s\right)  }{\cos s\sinh s+\sin s\cosh
s}\ ds,\label{fmo14}%
\end{equation}%
\begin{equation}
\gamma\left(  t\right)  =\int_{0}^{t}\left(  \alpha\left(  s\right)  g\left(
s\right)  -\frac{1}{2}\left(  1+\cos2s\right)  \alpha^{2}\left(  s\right)
\right)  \ ds.\label{fmo15}%
\end{equation}
The solution of the Schr\"{o}dinger equation (\ref{fmo5}) with the initial
condition (\ref{fmo2}) has the form%
\begin{equation}
\psi\left(  x,t\right)  =\int_{-\infty}^{\infty}G\left(  x,y,t\right)
\psi_{0}\left(  y\right)  \ dy\label{fmo16}%
\end{equation}
with the Green function given by%
\begin{equation}
G\left(  x,y,t\right)  =G_{0}\left(  x,y,t\right)  \ e^{i\left(  \alpha\left(
t\right)  x+\beta\left(  t\right)  y+\gamma\left(  t\right)  \right)
}.\label{fmo17}%
\end{equation}
Extensions for the $n$-dimensional case and to the corresponding
heat equation are obvious. The details are left to the reader.

\section{Expansion Formula for a Plane Wave}

Equations (\ref{gr9})--(\ref{gr10}) and (\ref{gr13}) imply the familiar
expansion formula of a plane wave in $\boldsymbol{R}^{n}$ in terms of the
hyperspherical harmonics%
\begin{equation}
e^{i\boldsymbol{x}\cdot\boldsymbol{x}^{\prime}}=rr^{\prime}\left(  \frac{2\pi
}{rr^{\prime}}\right)  ^{n/2}\sum_{K\nu}i^{K}\ Y_{K\nu}^{\ast}\left(
\Omega\right)  \ Y_{K\nu}\left(  \Omega^{\prime}\right)  \ J_{K+n/2-1}\left(
rr^{\prime}\right)  , \label{exp1}%
\end{equation}
where%
\begin{equation}
J_{\nu}\left(  z\right)  =\frac{\left(  z/2\right)  ^{\nu}}{\Gamma\left(
\nu+1\right)  }\ ~_{0}F_{1}\left(
\begin{array}
[c]{c}%
-\\
\nu+1
\end{array}
;\ -\frac{\left(  rr^{\prime}\right)  ^{2}}{4}\right)  \label{exp2}%
\end{equation}
is the Bessel function. Although expansion (\ref{exp1}) is well known in the
three dimensional case \cite{Flu}, we were not able to find it in the
literature for the (hyper)spherical system of coordinates in $\boldsymbol{R}%
^{n},$ corresponding to a general binary tree in
Vilenkin--Kuznetsov--Smorodinski\u{\i}'s graphical approach \cite{Ni:Su:Uv}.
Dick Askey has informed us that this formula follows immediately from the
Funk--Hecke theorem \cite{An:As:Ro} and a familiar integral giving the Bessel
functions as a Fourier transform. Bochner essentially has this in his book
\cite{Boch}; also see Claus M\"{u}ller's lectures on spherical harmonics
\cite{Mull} and \cite{Muller}; it is probably in some notes of Calderon
published in Argentina, but they are not widely available. Also see
\cite{AveryBook}, \cite{Avery}, and recent papers \cite{Bez:Dabr:Str} and
\cite{Bez:Str}.\smallskip

Let us consider a few examples.\smallskip

\noindent\textbf{Case }$n=1.\ $This is simply Euler's formula%
\begin{equation}
e^{i\varphi}=\cos\varphi+i\sin\varphi\label{exp2a}%
\end{equation}
in view of the familiar relations%
\begin{equation}
J_{-1/2}\left(  z\right)  =\sqrt{\frac{2}{\pi z}}\ \cos z,\qquad
J_{1/2}\left(  z\right)  =\sqrt{\frac{2}{\pi z}}\ \sin z. \label{exp2b}%
\end{equation}
%
%%%%%%%%%%%%%%%%%%%%%%%%%%%%%%%%%%%%%%%%%%%
%%%This WinEdt version of the figure 2%%%
\begin{figure}[ptbh]
\centering\scalebox{.65}{\includegraphics{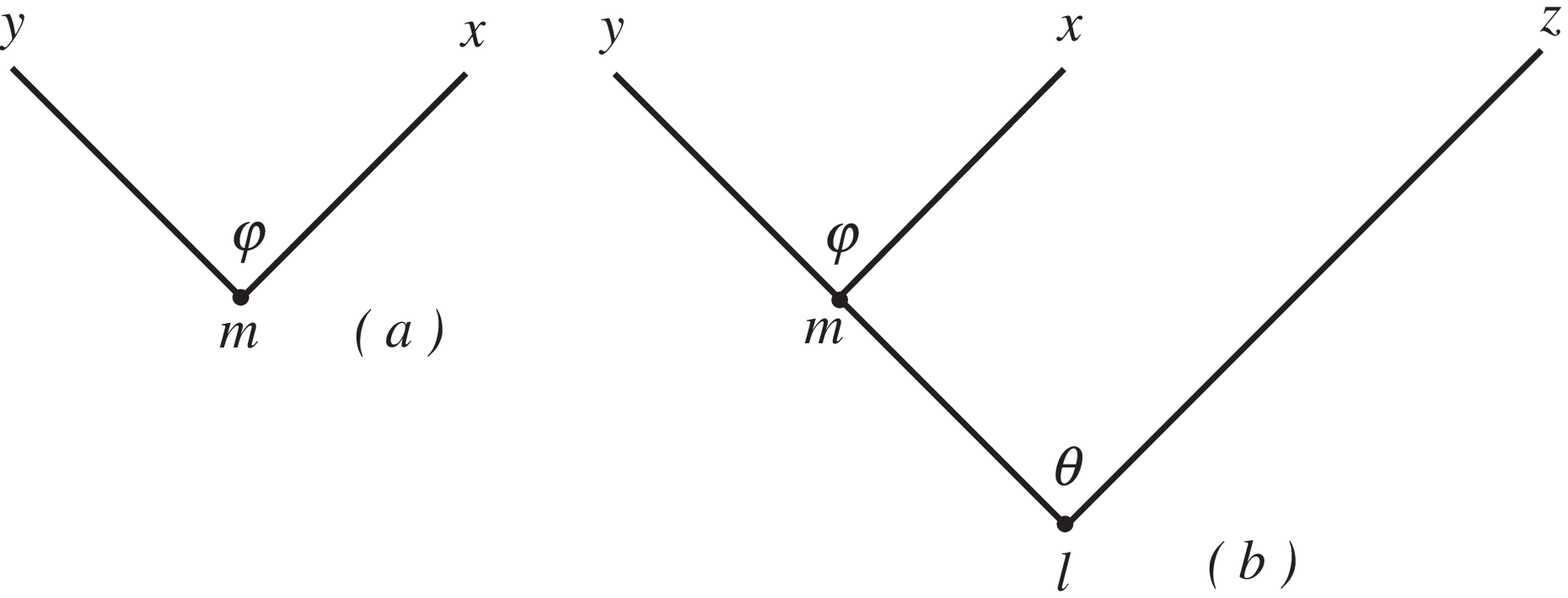}}\caption{Spherical
harmonics in $\boldsymbol{R}^{2}$ and $\boldsymbol{R}^{3}.$}%
\end{figure}
%%%%%%%%%%%%%%%%%%%%%%%%%%%%%%%%%%%%%%%%%%%

\noindent\textbf{Case }$n=2.\ \ $Let $x=r\cos\theta$ and $y=r\sin\theta;$ see
Figure~2(a). Then the expansion formula (\ref{exp1}) simplifies to%
\begin{equation}
e^{irr^{\prime}\cos\left(  \theta-\theta^{\prime}\right)  }=\sum_{m=-\infty
}^{\infty}i^{m}J_{m}\left(  rr^{\prime}\right)  \ e^{im\left(  \theta
-\theta^{\prime}\right)  }, \label{exp3}%
\end{equation}
or%
\begin{equation}
e^{iz\sin\varphi}=\sum_{m=-\infty}^{\infty}J_{m}\left(  z\right)
\ e^{im\varphi}. \label{exp3a}%
\end{equation}
This is a well known relation in the theory of Bessel functions; see, for
example, \cite{An:As:Ro} and \cite{Ni:Uv}.\smallskip

\noindent\textbf{Case }$n=3.\ \ $If $x=r\sin\theta\cos\varphi,$ $y=r\sin
\theta\sin\varphi,$ $z=r\cos\theta$ (Figure~2(b)), the expansion formula
(\ref{exp1}) takes the familiar form%
\begin{equation}
e^{i\boldsymbol{x}\cdot\boldsymbol{x}^{\prime}}=\frac{\left(  2\pi\right)
^{3/2}}{\sqrt{rr^{\prime}}}\ \sum_{l=0}^{\infty}\sum_{m=-l}^{l}i^{l}%
\ J_{l+1/2}\left(  rr^{\prime}\right)  \ Y_{lm}^{\ast}\left(  \theta
,\varphi\right)  Y_{lm}\left(  \theta^{\prime},\varphi^{\prime}\right)  ,
\label{exp4}%
\end{equation}
see \cite{Flu}, \cite{La:Lif}, and \cite{Ni:Uv} for more details.\smallskip
%%%%%%%%%%%%%%%%%%%%%%%%%%%%%%%%%%%%%%%%%%%
%%%This WinEdt version of the figure 3%%%
\begin{figure}[ptbh]
\centering\scalebox{.65}{\includegraphics{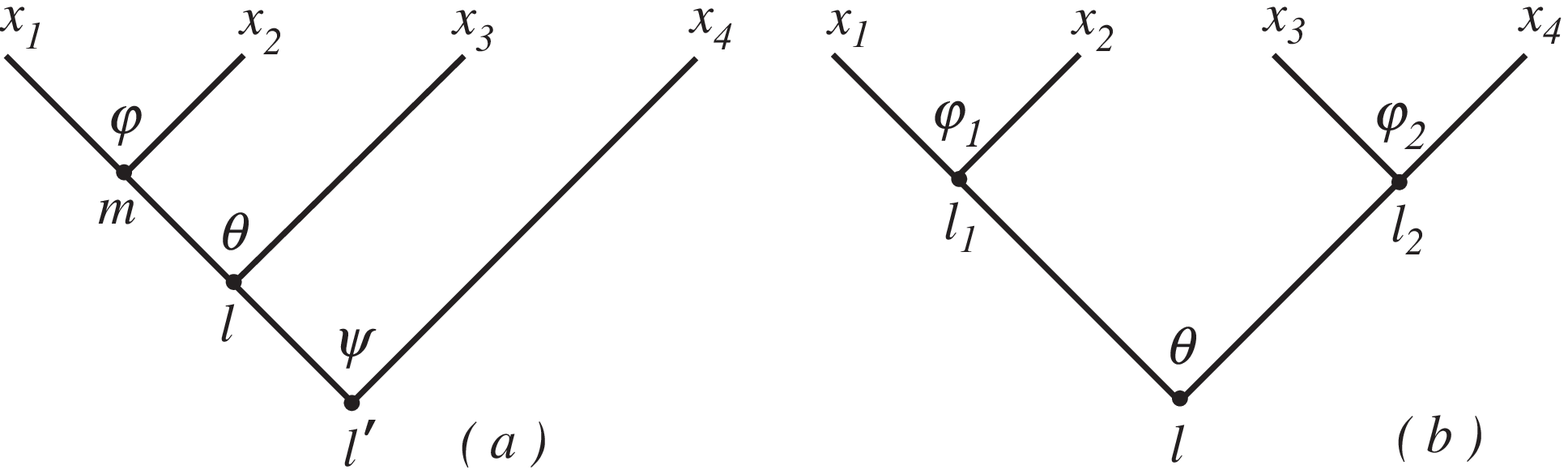}}\caption{Spherical
harmonics in $\boldsymbol{R}^{4}.$}%
\end{figure}
%%%%%%%%%%%%%%%%%%%%%%%%%%%%%%%%%%%%%%%%%%%

\noindent\textbf{Case }$n=4.\ $Two different system of the spherical
coordinates in $\boldsymbol{R}^{4}$ are%
\begin{align*}
x_{1} &  =r\sin\psi\sin\theta\sin\varphi,\\
x_{2} &  =r\sin\psi\sin\theta\cos\varphi,\\
x_{3} &  =r\sin\psi\cos\theta,\\
x_{4} &  =r\cos\psi
\end{align*}
and%
\begin{align*}
x_{1} &  =r\sin\theta\sin\varphi_{1},\\
x_{2} &  =r\sin\theta\cos\varphi_{1},\\
x_{3} &  =r\cos\theta\sin\varphi_{2},\\
x_{4} &  =r\cos\theta\cos\varphi_{2},
\end{align*}
see Figures~3(a)--(b), respectively. The corresponding spherical harmonics are
given by%
\begin{align}
&  Y_{l^{\prime}lm}\left(  \psi,\theta,\varphi\right)  =A\sin^{l}%
\psi\ P_{l^{\prime}-l}^{\left(  l+1/2,\ l+1/2\right)  }\left(  \cos
\psi\right)  \label{exp5}\\
&  \qquad\qquad\qquad\quad\times\sin^{\left\vert m\right\vert }\theta
\ P_{l-\left\vert m\right\vert }^{\left(  \left\vert m\right\vert
,\ \left\vert m\right\vert \right)  }\left(  \cos\theta\right)  \ e^{im\varphi
}\qquad\left(  l^{\prime}\geq l\geq\left\vert m\right\vert \right)  ,\nonumber
\end{align}
where%
\[
A=\frac{\sqrt{\left(  2l+1\right)  \left(  2l^{\prime}+2\right)  \left(
l-m\right)  !\left(  l+m\right)  !\left(  l^{\prime}-l\right)  !\left(
l^{\prime}+l+1\right)  !}}{\sqrt{\pi}\ 2^{l+\left\vert m\right\vert
-2}l!\Gamma\left(  l^{\prime}+3/2\right)  },
\]
and%
\begin{align}
&  Y_{l\ l_{1}l_{2}}\left(  \theta,\varphi_{1},\varphi_{2}\right)
=\frac{e^{i\left(  l_{1}\varphi_{1}+l_{2}\varphi_{2}\right)  }}{2\pi}%
\ N\sin^{\left\vert l_{1}\right\vert }\theta\ \cos^{\left\vert l_{2}%
\right\vert }\theta\ P_{\left(  l-\left\vert l_{1}\right\vert -\left\vert
l_{2}\right\vert \right)  /2}^{\left(  \left\vert l_{1}\right\vert
,\ \left\vert l_{2}\right\vert \right)  }\left(  \cos2\theta\right)
\label{exp6}\\
&  \qquad\qquad\quad\left(  l=\left\vert l_{1}\right\vert +\left\vert
l_{2}\right\vert ,\left\vert l_{1}\right\vert +\left\vert l_{2}\right\vert
+2,\left\vert l_{1}\right\vert +\left\vert l_{2}\right\vert +4,...\ .\right)
\nonumber
\end{align}
with%
\[
N=\sqrt{\frac{\left(  2l+2\right)  \left[  \left(  l-\left\vert l_{1}%
\right\vert -\left\vert l_{2}\right\vert \right)  /2\right]  !\left[  \left(
l+\left\vert l_{1}\right\vert +\left\vert l_{2}\right\vert \right)  /2\right]
!}{\left[  \left(  l+\left\vert l_{1}\right\vert -\left\vert l_{2}\right\vert
\right)  /2\right]  !\left[  \left(  l-\left\vert l_{1}\right\vert +\left\vert
l_{2}\right\vert \right)  /2\right]  !},}%
\]
respectively. Here $P_{n}^{\left(  \alpha,\ \beta\right)  }\left(  \xi\right)
$ are the Jacobi polynomials.\smallskip

The expansion formulas take the forms%
\begin{equation}
e^{i\boldsymbol{x}\cdot\boldsymbol{x}^{\prime}}=\frac{\left(  2\pi\right)
^{2}}{rr^{\prime}}\sum_{l^{\prime}=0}^{\infty}\sum_{l=0}^{l^{\prime}}%
\sum_{m=-l}^{l}i^{l^{\prime}}\ Y_{l^{\prime}lm}^{\ast}\left(  \psi
,\theta,\varphi\right)  \ Y_{l^{\prime}lm}^{\ast}\left(  \psi^{\prime}%
,\theta^{\prime},\varphi^{\prime}\right)  \ J_{l^{\prime}+1}\left(
rr^{\prime}\right)  \label{exp7}%
\end{equation}
and%
\begin{equation}
e^{i\boldsymbol{x}\cdot\boldsymbol{x}^{\prime}}=\frac{\left(  2\pi\right)
^{2}}{rr^{\prime}}\sum_{\left(  l-\left\vert l_{1}\right\vert -\left\vert
l_{2}\right\vert \right)  /2\geq0}\ \underset{l/2\geq\left(  \left\vert
l_{1}\right\vert +\left\vert l_{2}\right\vert \right)  /2\geq0}{\sum\sum}%
i^{l}\ Y_{l\ l_{1}l_{2}}^{\ast}\left(  \theta,\varphi_{1},\varphi_{2}\right)
\ Y_{l\ l_{1}l_{2}}\left(  \theta^{\prime},\varphi_{1}^{\prime},\varphi
_{2}^{\prime}\right)  \ J_{l+1}\left(  rr^{\prime}\right)  , \label{exp8}%
\end{equation}
respectively.\smallskip
%%%%%%%%%%%%%%%%%%%%%%%%%%%%%%%%%%%%%%%%%%%
%%%This WinEdt version of the figure 4%%%
\begin{figure}[ptbh]
\centering\scalebox{.65}{\includegraphics{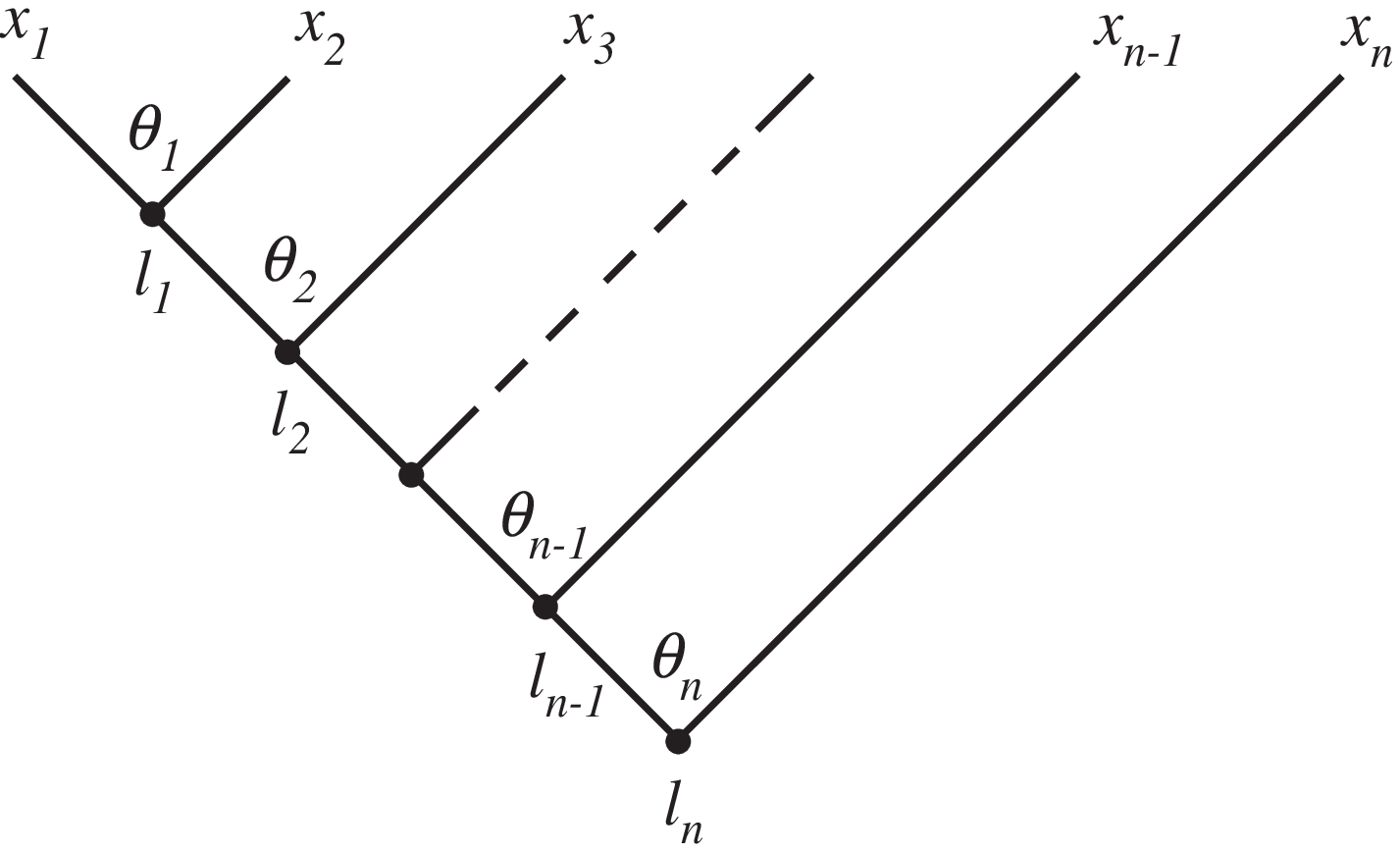}}\caption{Spherical
harmonics in $\boldsymbol{R}^{n}.$}%
\end{figure}
%%%%%%%%%%%%%%%%%%%%%%%%%%%%%%%%%%%%%%%%%%%

\noindent\textbf{The }$\boldsymbol{n}$-\textbf{dimensional case}$.\ $The
canonical system of hyperspherical coordinates in the Euclidean space
$\boldsymbol{R}^{n}$ can be set up as follows%
\begin{align*}
&  x_{1}=r\sin\theta_{n-1}\sin\theta_{n-2}\cdot\cdot\cdot\sin\theta_{2}%
\sin\theta_{1},\\
&  x_{2}=r\sin\theta_{n-1}\sin\theta_{n-2}\cdot\cdot\cdot\sin\theta_{2}%
\cos\theta_{1},\\
&  x_{3}=r\sin\theta_{n-1}\sin\theta_{n-2}\cdot\cdot\cdot\cos\theta_{2},\\
&  \quad.............................\\
&  x_{n-1}=r\sin\theta_{n-1}\cos\theta_{n-2},\\
&  x_{n}=r\cos\theta_{n-1}%
\end{align*}
with $0\leq\theta_{1}\leq2\pi,$ $0\leq\theta_{k}\leq\pi$ and $k=2,...\ ,n-1\ $%
(Figure$~$4); see \cite{An:As:Ro} and \cite{Ni:Su:Uv} for more
details.\smallskip

The hyperspherical harmonics have the form%
\begin{align}
&  Y_{l_{n-1}l_{n-2}...\ l_{1}}\left(  \theta_{n-1},\theta_{n-2}%
,...\ ,\theta_{1}\right)  =A\
%TCIMACRO{\dprod _{k=3}^{n-1}}%
%BeginExpansion
{\displaystyle\prod_{k=3}^{n-1}}
%EndExpansion
\sin^{l_{k-1}}\theta_{k}\ P_{l_{k+1}-l_{k}}^{\left(  2j_{k-1}+1,\ 2j_{k-1}%
+1\right)  }\left(  \theta_{k}\right) \label{exp9}\\
&  \qquad\qquad\qquad\qquad\qquad\qquad\qquad\times\sin^{\left\vert
l_{1}\right\vert }\theta_{2}\ P_{l_{2}-l_{1}}^{\left(  \left\vert
2j_{1}+1\right\vert ,\ \left\vert 2j_{1}+1\right\vert \right)  }\left(
\cos\theta_{2}\right)  \ e^{il_{1}\theta_{1}}\nonumber\\
&  \qquad\qquad\qquad\qquad\left(  l_{n-1}\geq l_{n-2}\geq...\ \geq l_{2}%
\geq\left\vert l_{1}\right\vert \right)  ,\nonumber
\end{align}
where $2j_{k}+1=l_{k}+\left(  k-1\right)  /2$ and $A$ is a normalizing
constant. The expansion formula is%
\begin{align}
e^{i\boldsymbol{x}\cdot\boldsymbol{x}^{\prime}}  &  =rr^{\prime}\left(
\frac{2\pi}{rr^{\prime}}\right)  ^{n/2}\sum_{l_{n-1}\geq l_{n-2}\geq...\ \geq
l_{2}\geq\left\vert l_{1}\right\vert }i^{l_{n-1}}\ J_{l_{n-1}+n/2-1}\left(
rr^{\prime}\right) \label{exp10}\\
&  \times Y_{l_{n-1}l_{n-2}...\ l_{1}}^{\ast}\left(  \theta_{n-1},\theta
_{n-2},...\ ,\theta_{1}\right)  \ Y_{l_{n-1}l_{n-2}...\ l_{1}}\left(
\theta_{n-1},\theta_{n-2},...\ ,\theta_{1}\right)  .\nonumber
\end{align}
We leave the details to the reader.

\section{The Schr\"{o}dinger Equation on the Group $SU\left(  1,1\right)  $}

Equation (\ref{i7}) admits the following algebraic generalization. Let
$J_{\pm}$ and $J_{0}$ be infinitesimal operators of the Lie algebra $SU\left(
1,1\right)  $ and $\left\vert j,m\right\rangle =\psi_{jm}\left(  x\right)  $
be a basis of the irreducible representation $\mathcal{D}_{+}^{j}$ belonging
to the discrete positive series in certain Hilbert space \cite{Bargmann47}. We
may call the equation%
\begin{equation}
i\frac{\partial\psi}{\partial t}=H\left(  t\right)  \psi\label{su1}%
\end{equation}
with the Hamiltonian of the form%
\begin{equation}
H\left(  t\right)  =\omega\left(  t\right)  J_{0}+\delta\left(  t\right)
J_{+}+\delta^{\ast}\left(  t\right)  J_{-}, \label{su2}%
\end{equation}
where $\omega\left(  t\right)  $ is a real valued function of time and
$\delta\left(  t\right)  $ is a complex valued function of time, the
time-dependent Schr\"{o}dinger equation on the group $SU\left(  1,1\right)
.$\smallskip

The eigenfunction expansion%
\begin{equation}
\psi\left(  x,t\right)  =\sum_{m=j+1}^{\infty}c_{m}\left(  t\right)
\ \psi_{jm}\left(  x\right)  \label{su3}%
\end{equation}
results in the system%
\begin{align}
i\frac{dc_{m}\left(  t\right)  }{dt} &  =\omega\left(  t\right)
m\ c_{m}\left(  t\right)  +\delta\left(  t\right)  \sqrt{\left(  m-j-1\right)
\left(  m+j\right)  }\ c_{m-1}\left(  t\right)  \label{su4}\\
&  \quad+\delta^{\ast}\left(  t\right)  \sqrt{\left(  m+j+1\right)  \left(
m-j\right)  }\ c_{m+1}\left(  t\right)  \nonumber
\end{align}
with $m=j+1,j+2,...\ $ similar to (\ref{separ3}). When $\delta\left(
t\right)  =e^{-i\omega t}$ and $\omega=$cons$\tan$t$,$ the same considerations
as in sections~3 and 6 give us the solution of the corresponding Cauchy
initial value problem in the form of the expansion (\ref{su3}) with the
time-dependent coefficients
\begin{equation}
c_{m}\left(  t\right)  =e^{-i\omega mt}\sum_{m^{\prime}=j+1}^{\infty
}i^{m^{\prime}-m}v_{m^{\prime}m}^{j}\left(  2t\right)  \ \left(
\psi_{jm^{\prime}},\ \psi_{0}\right)  ,\label{su5}%
\end{equation}
where $v_{m^{\prime}m}^{j}\left(  2t\right)  $ are the Bargmann functions
(\ref{barg2}) and
\begin{equation}
\left(  \varphi,\ \chi\right)  =\int_{\text{Supp\ }\mu}\varphi^{\ast}\left(
x\right)  \chi\left(  x\right)  \ d\mu\left(  x\right)  \label{su6}%
\end{equation}
is the inner product in the Hilbert space under consideration. A formal
substitution of (\ref{su5}) into (\ref{su3}) results in%
\begin{equation}
\psi\left(  x,t\right)  =\int_{\text{Supp\ }\mu}G\left(  x,y,t\right)
\ \psi_{0}\left(  y\right)  \ d\mu\left(  y\right)  \label{su7}%
\end{equation}
with%
\begin{equation}
G\left(  x,y,t\right)  =\sum_{m=j+1}^{\infty}\sum_{m^{\prime}=j+1}^{\infty
}e^{-i\omega mt}\ i^{m^{\prime}-m}\ v_{m^{\prime}m}^{j}\left(  2t\right)
\ \psi_{jm}\left(  x\right)  \psi_{jm^{\prime}}\left(  y\right)  .\label{su8}%
\end{equation}
The original Cauchy initial value problem (\ref{i7})--(\ref{i9}) provides an
explicit model of the abstract Hilbert space. Another realization is a
quasipotential model of the relativistic oscillator; see, for example,
\cite{Log:Tavk}, \cite{At:Mir:Nag}, and \cite{At:SusCH}. We shall discuss this
model and its generalization in the next two sections. Propagators for the
difference models of the simple harmonic oscillator \cite{At:SusOsc} can be
derived in a similar fashion. The details are left to the reader.

\section{Propagator for the Relativistic Oscillator}

For the consistent three-dimensional description of a relativistic
two-particle system in quantum field theory a quasipotential approach had been
formulated and, in the framework of this approach, some relativistic
generalizations of the exactly solvable problems of quantum mechanics have
been considered; see \cite{Log:Tavk}, \cite{At:Mir:Nag}, and references
therein for more information. We shall discuss here a model of the (simple)
relativistic oscillator described by the following Hamiltonian operator%
\begin{equation}
H_{0}=mc^{2}\cosh\left(  i\lambda\partial_{x}\right)  +\frac{1}{2}m\omega
^{2}x\left(  x+i\lambda\right)  \exp\left(  i\lambda\partial_{x}\right)  ,
\label{sro1}%
\end{equation}
where $\lambda=\hslash/mc$ is the Compton wave length, $\partial_{x}%
=\partial/\partial x$ and $\exp\left(  \alpha\partial_{x}\right)  f\left(
x\right)  =f\left(  x+\alpha\right)  $ is the shift operator. The
square-integrable solutions of the stationary Schr\"{o}dinger equation%
\begin{equation}
H_{0}\Psi_{n}\left(  x\right)  =E_{n}\Psi_{n}\left(  x\right)  \qquad\left(
-\infty<x<\infty\right)  \label{sro2}%
\end{equation}
on the real line, which correspond to the discrete energy levels%
\begin{equation}
E_{n}=\hslash\omega\left(  n+\nu\right)  ,\quad\nu=\frac{1}{2}+\sqrt{\frac
{1}{4}+\left(  \frac{c}{\lambda\omega}\right)  ^{2}}\qquad\left(
n=0,1,2,...\ \right)  , \label{sro2a}%
\end{equation}
can be found in terms of the special Meixner--Pollaczek polynomials
\cite{At:SusCH} as follows%
\begin{equation}
\Psi_{n}\left(  x\right)  =2^{\nu}\sqrt{\frac{n!}{2\pi\lambda\Gamma\left(
n+2\nu\right)  }}\ \left(  \nu\left(  \nu-1\right)  \right)  ^{-ix/2\lambda
}\ \Gamma\left(  \nu+ix/\lambda\right)  \ P_{n}^{\nu}\left(  x/\lambda
,\ \pi/2\right)  . \label{sro3}%
\end{equation}

On the other hand, solution of the Cauchy initial value problem for the
time-dependent Schr\"{o}dinger equation%
\begin{equation}
i\hslash\frac{\partial\psi}{\partial t}=H_{0}\psi\label{sro4}%
\end{equation}
with the Hamiltonian (\ref{sro1}) subject to the initial condition%
\begin{equation}
\left.  \psi\left(  x,t\right)  \right\vert _{t=0}=\psi_{0}\left(  x\right)
\in\mathcal{L}^{2}\left(  -\infty,\infty\right)  \label{sro5}%
\end{equation}
has the form%
\begin{align}
\psi\left(  x,t\right)   &  =\sum_{n=0}^{\infty}e^{-i\left(  E_{n}\ t\right)
/\hslash}\ \Psi_{n}\left(  x\right)  \ \int_{-\infty}^{\infty}\Psi_{n}^{\ast
}\left(  y\right)  \psi_{0}\left(  y\right)  \ dy\label{sro6}\\
&  =\int_{-\infty}^{\infty}G_{0}\left(  x,y,t\right)  \ \psi_{0}\left(
y\right)  \ dy,\nonumber
\end{align}
where the Green function (or Feynman's propagator)
\begin{equation}
G_{0}\left(  x,y,t\right)  =\sum_{n=0}^{\infty}e^{-i\omega\left(
n+\nu\right)  t}\ \Psi_{n}\left(  x\right)  \ \Psi_{n}^{\ast}\left(  y\right)
\label{sro7}%
\end{equation}
can be found in a closed form as follows%
\begin{align}
G_{0}\left(  x,y,t\right)   &  =\frac{1}{2\pi\lambda}\left(  \frac{c}%
{\lambda\omega}\right)  ^{i\left(  y-x\right)  /\lambda}\ \left(  i\sin\left(
\omega t/2\right)  \right)  ^{i\left(  y-x\right)  /\lambda}\left(
\cos\left(  \omega t/2\right)  \right)  ^{-i\left(  x+y\right)  /\lambda
}\label{sro8}\\
&  \qquad\quad\times\ \Gamma\left(  i\left(  x-y\right)  /\lambda\right)
\ _{2}F_{1}\left(
\begin{array}
[c]{c}%
\nu-ix/\lambda,\ 1-\nu-ix/\lambda\\
1+i\left(  y-x\right)  /\lambda
\end{array}
;\ \sin^{2}\left(  \omega t/2\right)  \right) \nonumber\\
&  \quad+\frac{1}{2\pi\lambda}\left(  \frac{c}{\lambda\omega}\right)
^{i\left(  y-x\right)  /\lambda}\ \left(  i\sin\left(  \omega t/2\right)
\right)  ^{i\left(  x-y\right)  /\lambda}\left(  \cos\left(  \omega
t/2\right)  \right)  ^{-i\left(  x+y\right)  /\lambda}\nonumber\\
&  \quad\quad\quad\quad\times\ \Gamma\left(  i\left(  y-x\right)
/\lambda\right)  \ \frac{\Gamma\left(  \nu+ix/\lambda\right)  \Gamma\left(
\nu-iy/\lambda\right)  }{\Gamma\left(  \nu-ix/\lambda\right)  \Gamma\left(
\nu+iy/\lambda\right)  }\nonumber\\
&  \qquad\qquad\qquad\qquad\quad\times\ _{2}F_{1}\left(
\begin{array}
[c]{c}%
\nu-iy/\lambda,\ 1-\nu-iy/\lambda\\
1+i\left(  x-y\right)  /\lambda
\end{array}
;\ \sin^{2}\left(  \omega t/2\right)  \right)  .\nonumber
\end{align}
We have used here the Poisson kernel for the Meixner--Pollaczek polynomials
(\ref{pollaczek5}) in order to sum the series and a transformation formula for
the analytic continuation of the hypergeometric function \cite{An:As:Ro} and
\cite{Ni:Uv}.

\section{A Modified Relativistic Oscillator}

As is known \cite{At:Mir:Nag} a dynamical symmetry group for the relativistic
oscillator with the Hamiltonian (\ref{sro1}) is the group $SU\left(
1,1\right)  $ (or isomorphic groups $SO\left(  2,1\right)  \sim Sp\left(
2,\boldsymbol{R}\right)  $ $\sim SL\left(  2,\boldsymbol{R}\right)  $), whose
generators are realized as the difference operators%
\begin{equation}
K_{0}=\frac{1}{\hslash\omega}H_{0},\qquad K_{\pm}=\frac{x}{\lambda}\pm
iK_{0}\mp\frac{ic}{\lambda\omega}\exp\left(  -i\lambda\partial_{x}\right)
\label{mro1}%
\end{equation}
acting on the eigenfunctions of the Hamiltonian (\ref{sro1}) as follows%
\begin{align}
K_{+}\Psi_{n}\left(  x\right)   &  =\sqrt{\left(  n+1\right)  \left(
n+2\nu\right)  }\ \Psi_{n+1}\left(  x\right)  ,\label{mro2}\\
K_{-}\Psi_{n}\left(  x\right)   &  =\sqrt{n\left(  n+2\nu-1\right)  }%
\ \Psi_{n-1}\left(  x\right)  ,\nonumber\\
K_{0}\Psi_{n}\left(  x\right)   &  =\left(  n+\nu\right)  \ \Psi_{n}\left(
x\right)  .\nonumber
\end{align}
The wavefunctions $\Psi_{n}\left(  x\right)  =\psi_{jm}$ with $n=m-j-1$ and
$\nu=j+1$ form the basis for the infinite-dimensional irreducible unitary
representations of the discrete positive series $\mathcal{D}_{+}^{j}$ of the
universal covering group $\widetilde{SU}\left(  1,1\right)  ;$ see equations
(\ref{harm14}).\smallskip

In this section, we shall solve the time-dependent Schr\"{o}dinger equation%
\begin{equation}
i\hslash\frac{\partial\psi}{\partial t}=H\left(  t\right)  \psi\label{mro3}%
\end{equation}
with a modified Hamiltonian of the form%
\begin{equation}
H\left(  t\right)  =H_{0}+\hslash\left(  \delta\left(  t\right)  K_{+}%
+\delta^{\ast}\left(  t\right)  K_{-}\right)  ,\qquad\delta\left(  t\right)
=e^{-i\omega t} \label{mro4}%
\end{equation}
subject to the initial condition (\ref{sro5}). The eigenfunction expansion%
\begin{equation}
\psi\left(  x,t\right)  =\sum_{m=j+1}^{\infty}c_{m}\left(  t\right)
\ \psi_{jm}\left(  x\right)  \label{mro5}%
\end{equation}
leads to the familiar system (\ref{separ3}), whose solutions in terms of the
Bargmann functions are%
\begin{equation}
c_{m}\left(  t\right)  =e^{-i\omega mt}\sum_{m^{\prime}=j+1}^{\infty
}i^{m^{\prime}-m}v_{m^{\prime}m}^{j}\left(  2t\right)  \ \int_{-\infty
}^{\infty}\psi_{jm^{\prime}}^{\ast}\left(  y\right)  \psi_{0}\left(  y\right)
\ dy. \label{mro6}%
\end{equation}
Thus%
\begin{equation}
\psi\left(  x,t\right)  =\int_{-\infty}^{\infty}G\left(  x,y,t\right)
\ \psi_{0}\left(  y\right)  \ dy, \label{mro7}%
\end{equation}
where the Green function is given by%
\begin{equation}
G\left(  x,y,t\right)  =\sum_{m=j+1}^{\infty}\sum_{m^{\prime}=j+1}^{\infty
}e^{-i\omega mt}\ i^{m^{\prime}-m}\ v_{m^{\prime}m}^{j}\left(  2t\right)
\ \psi_{jm}\left(  x\right)  \psi_{jm^{\prime}}^{\ast}\left(  y\right)  .
\label{mro8}%
\end{equation}
This multiple series can be simplified to two single sums. Indeed, rewriting
the wave functions (\ref{sro3}) in terms of the hypergeometric function%
\begin{align}
\psi_{jm}\left(  x\right)   &  =\frac{2^{j+1}}{\sqrt{2\pi\lambda}}%
\frac{\left(  j\left(  j+1\right)  \right)  ^{-ix/2\lambda}\Gamma\left(
j+1+ix/\lambda\right)  }{\Gamma\left(  2j+2\right)  }\sqrt{\frac{\left(
m+j\right)  !}{\left(  m-j-1\right)  !}}\label{mro8a}\\
&  \quad\quad\times\left(  -i\right)  ^{m-j-1}\ _{2}F_{1}\left(
\begin{array}
[c]{c}%
-m+j+1,\ j+1-ix/\lambda\\
2j+2
\end{array}
;\ 2\right) \nonumber
\end{align}
by (\ref{pollaczek1}) and (\ref{meixner1}) and then using the Meixner
generating relation (\ref{bilinearmeixner}), we have%
\begin{align}
&  \sum_{m^{\prime}=j+1}^{\infty}i^{m^{\prime}-j-1}\ v_{m^{\prime}m}%
^{j}\left(  2t\right)  \ \psi_{jm^{\prime}}^{\ast}\left(  y\right)
\label{mro9}\\
&  \quad=\frac{2^{j+1}}{\sqrt{2\pi\lambda}}\frac{\left(  j\left(  j+1\right)
\right)  ^{iy/\lambda}\Gamma\left(  j+1-iy/\lambda\right)  }{\Gamma\left(
2j+2\right)  }\ e^{-2ity/\lambda}\nonumber\\
&  \qquad\times\left(  -1\right)  ^{m-j-1}\sqrt{\frac{\left(  m+j\right)
!}{\left(  m-j-1\right)  !}}\ _{2}F_{1}\left(
\begin{array}
[c]{c}%
-m+j+1,\ j+1+iy/\lambda\\
2j+2
\end{array}
;\ 2\right) \nonumber\\
&  \quad=i^{m-j-1}\ e^{-2ity/\lambda}\ \psi_{jm}^{\ast}\left(  y\right)
.\nonumber
\end{align}
Thus, say for a pure imaginary time,%
\begin{align}
G\left(  x,y,t\right)   &  =\sum_{m=j+1}^{\infty}e^{-i\omega mt}\ \left(
-i\right)  ^{m-j-1}\ \psi_{jm}\left(  x\right) \label{mro10}\\
&  \quad\times\sum_{m^{\prime}=j+1}^{\infty}i^{m^{\prime}-j-1}\ v_{m^{\prime
}m}^{j}\left(  2t\right)  \ \psi_{jm^{\prime}}^{\ast}\left(  y\right)
\nonumber\\
&  =\frac{1}{2\pi\lambda}\left(  j\left(  j+1\right)  \right)  ^{i\left(
y-x\right)  /2\lambda}\frac{\Gamma\left(  j+1+ix/\lambda\right)  \Gamma\left(
j+1-iy/\lambda\right)  }{\Gamma\left(  2j+2\right)  }\nonumber\\
&  \qquad\times e^{-2ity/\lambda}\ \left(  \cos\left(  \omega t/2\right)
\right)  ^{-2j-2}\left(  i\tan\left(  \omega t/2\right)  \right)  ^{i\left(
y-x\right)  /\lambda}\nonumber\\
&  \qquad\quad\times\ _{2}F_{1}\left(
\begin{array}
[c]{c}%
j+1-ix/\lambda,\ j+1+iy/\lambda\\
2j+2
\end{array}
;\ \frac{1}{\cos^{2}\left(  \omega t/2\right)  }\right)  .\nonumber
\end{align}
Here we use a transformation formula for the analytic continuation of the
hypergeometric function \cite{An:As:Ro} and \cite{Ni:Uv} in order to obtain
the final result as follows%
\begin{align}
G\left(  x,y,t\right)   &  =\frac{1}{2\pi\lambda}\left(  \frac{c}%
{\lambda\omega}\right)  ^{i\left(  y-x\right)  /\lambda}\ e^{-2ity/\lambda
}\ \left(  i\sin\left(  \omega t/2\right)  \right)  ^{i\left(  y-x\right)
/\lambda}\ \left(  \cos\left(  \omega t/2\right)  \right)  ^{-i\left(
x+y\right)  /\lambda}\label{mro11}\\
&  \qquad\times\Gamma\left(  i\left(  x-y\right)  /\lambda\right)  \ _{2}%
F_{1}\left(
\begin{array}
[c]{c}%
\nu-ix/\lambda,\ 1-\nu-ix/\lambda\\
1+i\left(  y-x\right)  /\lambda
\end{array}
;\ \sin^{2}\left(  \omega t/2\right)  \right) \nonumber\\
&  \quad+\frac{1}{2\pi\lambda}\left(  \frac{c}{\lambda\omega}\right)
^{i\left(  y-x\right)  /\lambda}\ e^{-2ity/\lambda}\ \left(  i\sin\left(
\omega t/2\right)  \right)  ^{i\left(  x-y\right)  /\lambda}\ \left(
\cos\left(  \omega t/2\right)  \right)  ^{i\left(  x+y\right)  /\lambda
}\nonumber\\
&  \qquad\quad\times\Gamma\left(  i\left(  y-x\right)  /\lambda\right)
\frac{\Gamma\left(  \nu+ix/\lambda\right)  \Gamma\left(  \nu-iy/\lambda
\right)  }{\Gamma\left(  \nu-ix/\lambda\right)  \Gamma\left(  \nu
+iy/\lambda\right)  }\nonumber\\
&  \qquad\quad\quad\qquad\qquad\quad\times\ _{2}F_{1}\left(
\begin{array}
[c]{c}%
\nu+ix/\lambda,\ 1-\nu+ix/\lambda\\
1+i\left(  x-y\right)  /\lambda
\end{array}
;\ \sin^{2}\left(  \omega t/2\right)  \right)  .\nonumber
\end{align}
Comparing expressions (\ref{sro8}) and (\ref{mro11}) for two propagators, one
can see that%
\begin{equation}
G\left(  x,y,t\right)  =G_{0}\left(  x,y,t\right)  \ e^{-2ity/\lambda}.
\label{mro12}%
\end{equation}
It also follows from (\ref{sro7}), (\ref{mro8a}), (\ref{mro9}), and
(\ref{mro10}). The details are left to the reader.

\section{A System of Ordinary Differential Equations}

Consider the following infinite system of ordinary differential equations%
\begin{equation}
i\frac{du_{n}\left(  t\right)  }{dt}=\left(  n+1\right)  u_{n+1}\left(
t\right)  -2\left(  n+\lambda\right)  \cos\varphi\ u_{n}\left(  t\right)
+\left(  n+2\lambda-1\right)  u_{n-1}\left(  t\right)  , \label{sys1}%
\end{equation}
with $v_{-1}\left(  t\right)  =0$ $\left(  n=0,1,2,...\ \right)  $ subject to
the initial conditions%
\begin{equation}
u_{n}\left(  0\right)  =u_{n}^{0}. \label{sys0}%
\end{equation}
Looking for a particular solution in the form%
\begin{equation}
u_{n}\left(  t\right)  =e^{-2ixt\sin\varphi}\ P_{n}^{\lambda}\left(
x,\varphi\right)  , \label{sys2}%
\end{equation}
one gets the three term recurrence relation (\ref{pollaczek2}) for the
Meixner--Pollaczek polynomials $P_{n}^{\lambda}\left(  x,\varphi\right)  .$
The particular solution that satisfies the initial conditions $v_{nm}\left(
0\right)  =$ $\delta_{nm}$ is given in the integral form%
\begin{equation}
u_{nm}\left(  t\right)  =\frac{1}{d_{m}^{2}}\int_{-\infty}^{\infty
}e^{-2ixt\sin\varphi}\ P_{n}^{\lambda}\left(  x,\varphi\right)  \ P_{m}%
^{\lambda}\left(  x,\varphi\right)  \ \rho\left(  x\right)  dx, \label{sys3}%
\end{equation}
where $\rho\left(  x\right)  $and $d_{m}^{2}$ are the weight function and the
squared norm for the Meixner--Pollaczek polynomials $P_{n}^{\lambda}\left(
x,\varphi\right)  ,$ respectively; see (\ref{pollaczek3})--(\ref{pollaczek4}).
The solution of the initial value problem is%
\begin{equation}
u_{n}\left(  t\right)  =\sum_{m=0}^{\infty}u_{m}^{0}\ u_{nm}\left(  t\right)
. \label{sys4}%
\end{equation}
The last integral can be evaluated as a single sum with the help of Meixner's
generating relation (\ref{bilinearmeixner}). The final result is%
\begin{align}
u_{nm}\left(  t\right)   &  =\frac{\left(  2\lambda\right)  _{n}}{n!}%
\frac{e^{i\pi\lambda}\left(  \sin\varphi\right)  ^{2\lambda}\ \left(
\sinh\left(  t\sin\varphi\right)  \right)  ^{n+m}}{\left(  \cos\varphi
\sinh\left(  t\sin\varphi\right)  +i\sin\varphi\cosh\left(  t\sin
\varphi\right)  \right)  ^{n+m+2\lambda}}\label{sys5}\\
&  \qquad\times\ \ _{2}F_{1}\left(
\begin{array}
[c]{c}%
-n,\ -m\\
2\lambda
\end{array}
;\ -\left(  \frac{\sin\varphi}{\sinh\left(  t\sin\varphi\right)  }\right)
^{2}\right)  .\nonumber
\end{align}
This is an extention of (\ref{sol1})--(\ref{sol4}) where
$\varphi=\pi /2.$\smallskip

\noindent\textbf{Acknowledgment.\/} This paper is written as a part of the
summer 2007 program on analysis of the Mathematical and Theoretical Biology
Institute (MTBI) at Arizona State University. The MTBI/SUMS undergraduate
research program is supported by The National Science Foundation
(DMS--0502349), The National Security Agency (DOD--H982300710096), The Sloan
Foundation, and Arizona State University. The authors are grateful to
Professor Carlos Castillo-Ch\'{a}vez for support and reference
\cite{Bet:Cin:Kai:Cas}. We thank Professors Dick Askey, Slim Ibrahim, Erik
Koelink, Hunk Kuiper, Svetlana Roudenko, and Andreas Ruffing for valuable
comments. One of us (Maria Meiler) thanks MTBI/SUMS and the Department of
Mathematics and Statistics at Arizona State University for hospitality.

\section{Appendix. Another Integral Evaluation}

The following integral%
\begin{align}
&  \int_{0}^{\infty}\left.  e^{-\lambda z}z^{\gamma-1}\ \ _{1}F_{1}\left(
\begin{array}
[c]{c}%
\alpha\\
\gamma
\end{array}
;\ kz\right)  \ _{1}F_{1}\left(
\begin{array}
[c]{c}%
\alpha^{\prime}\\
\gamma
\end{array}
;\ k^{\prime}z\right)  \right.  \ dz\label{iap1}\\
&  \quad=\Gamma\left(  \gamma\right)  \lambda^{\alpha+\alpha^{\prime}-\gamma
}\left(  \lambda-k\right)  ^{-\alpha}\left(  \lambda-k^{\prime}\right)
^{-\alpha^{\prime}}\ _{2}F_{1}\left(
\begin{array}
[c]{c}%
\alpha,\ \alpha^{\prime}\\
\gamma
\end{array}
;\ \frac{kk^{\prime}}{\left(  \lambda-k\right)  \left(  \lambda-k^{\prime
}\right)  }\right) \nonumber
\end{align}
is evaluated in \cite{Gordon} and \cite{La:Lif}. When $\alpha^{\prime}%
=\alpha,$ replace $\lambda=\mu\alpha,$ $z=x/\alpha$ and take the limit
$\alpha=n\rightarrow\infty$ with the help of%
\begin{align}
&  \lim_{\alpha\rightarrow\infty}\ _{1}F_{1}\left(
\begin{array}
[c]{c}%
\alpha\\
\gamma
\end{array}
;\ \frac{kx}{\alpha}\right)  =~_{0}F_{1}\left(
\begin{array}
[c]{c}%
-\\
\gamma
\end{array}
;\ kx\right)  ,\label{iap2}\\
&  \lim_{\alpha\rightarrow\infty}\ _{2}F_{1}\left(
\begin{array}
[c]{c}%
\alpha,\ \alpha\\
\gamma
\end{array}
;\ \frac{kk^{\prime}}{\left(  \mu\alpha-k\right)  \left(  \mu\alpha-k^{\prime
}\right)  }\right)  =~_{0}F_{1}\left(
\begin{array}
[c]{c}%
-\\
\gamma
\end{array}
;\ \frac{kk^{\prime}}{\mu^{2}}\right)  ,\label{iap3}\\
&  \lim_{n\rightarrow\infty}\left(  1+\frac{\beta}{n}\right)  ^{n}=e^{\beta}.
\label{iap4}%
\end{align}
The end result%
\begin{align}
&  \int_{0}^{\infty}e^{-\mu x}\ x^{\gamma-1}\ \ _{0}F_{1}\left(
\begin{array}
[c]{c}%
-\\
\gamma
\end{array}
;\ kx\right)  \ _{0}F_{1}\left(
\begin{array}
[c]{c}%
-\\
\gamma
\end{array}
;\ k^{\prime}x\right)  \ dx\label{iap5}\\
&  \qquad=\Gamma\left(  \gamma\right)  \mu^{-\gamma}e^{\left(  k+k^{\prime
}\right)  /\mu}\ _{0}F_{1}\left(
\begin{array}
[c]{c}%
-\\
\gamma
\end{array}
;\ \frac{kk^{\prime}}{\mu^{2}}\right) \nonumber
\end{align}
is required in section~8 of this paper.

\end{document}